\documentclass[journal]{IEEEtran}

\ifCLASSINFOpdf
\else
   \usepackage[dvips]{graphicx}
\fi
\usepackage{url}

\usepackage{graphicx}
\usepackage{amsmath,amssymb,amsfonts}

\usepackage[linesnumbered,boxed,ruled,commentsnumbered]{algorithm2e}

\usepackage{cite}
\usepackage{subfigure}

\usepackage{color}

\begin{document}

\title{ Deterministic Pilot Design and Channel Estimation for Downlink Massive MIMO-OTFS Systems in Presence of the Fractional Doppler
}

\author{Ding Shi, \IEEEmembership{Graduate Student Member, IEEE}, Wenjin Wang, \IEEEmembership{Member, IEEE}, Li You, \IEEEmembership{Member, IEEE}, \\Xiaohang Song, \IEEEmembership{Member, IEEE}, Yi Hong, \IEEEmembership{Senior Member, IEEE}, Xiqi Gao, \IEEEmembership{Fellow, IEEE}, \\ and Gerhard Fettweis, \IEEEmembership{Fellow, IEEE}
	
\thanks{
This work was supported in part by the National Key Research and Development Program of China under Grant 2018YFB1801103; in part by the Jiangsu Province Basic Research Project under Grant BK20192002; in part by the National Natural Science Foundation of China under Grants 61761136016, 61631018, and 61801114; in part by the Fundamental Research Funds for the Central Universities; and in part by the German Science Foundation (DFG) through the Project Large-Scale and Hierarchical Bayesian Inference for Future Mobile Communication Networks under Grant 392016367.

D. Shi, W. J. Wang, L. You and X. Q. Gao are with the National Mobile Communications Research Laboratory, Southeast University, Nanjing 210096, China (e-mail: shiding@seu.edu.cn; wangwj@seu.edu.cn; liyou@seu.edu.cn; xqgao@seu.edu.cn).
	
X. H. Song and G. Fettweis are with the Vodafone Chair Mobile Communications Systems, Technische Universit{\"a}t Dresden, 01062 Dresden, Germany (e-mail: xiaohang.song@tu-dresden.de; gerhard.fettweis@tu-dresden.de).

Y. Hong is with the ECSE Department, Monash University, Clayton, VIC 3800, Australia (e-mail: Yi.Hong@monash.edu).
}

}

\maketitle

\begin{abstract}
Although the combination of the orthogonal time frequency space (OTFS) modulation and the massive multiple-input multiple-output (MIMO) technology can make communication systems perform better in high-mobility scenarios, there are still many challenges in downlink channel estimation owing to inaccurate modeling and high pilot overhead in practical systems.
In this paper, we propose a channel state information (CSI) acquisition scheme for downlink massive MIMO-OTFS in presence of the fractional Doppler, including deterministic pilot design and channel estimation algorithm. First, we analyze the input-output relationship of the single-input single-output (SISO) OTFS based on the orthogonal frequency division multiplexing (OFDM) modem and extend it to massive MIMO-OTFS. Moreover, we formulate an accurate model for the practical system in which the fractional Doppler is considered and the influence of subpaths is revealed. A deterministic pilot design is then proposed based on the model and the structure of the pilot matrix to reduce pilot overhead and save memory consumption.
Since channel geometry changes very slowly relative to the communication timescale, we put forward a modified sensing matrix based channel estimation (MSMCE) algorithm to acquire the downlink CSI. Simulation results demonstrate that the proposed downlink CSI acquisition scheme has significant advantages over traditional algorithms.
\end{abstract}


\IEEEpeerreviewmaketitle

\section{Introduction}

\IEEEPARstart{T}{o} meet the needs of future wireless communications, especially various new services and scenarios that are continually  emerging, the fifth-generation (5G) mobile systems come into being. One objective of 5G is to achieve reliable communication for the scenarios with high Doppler spread, such as high-mobility scenarios \cite{7383229,7833456}. Orthogonal frequency division multiplexing (OFDM) technology has been widely used to combat the inter-symbol interference (ISI) in time-invariant channels \cite{1096357,1266914,1033009,1190738}. However, when there is a high Doppler spread in time-variant channels, which would lead to severe inter-carrier interference (ICI), the performance of OFDM degrades significantly\cite{1638663}.

To cope with this problem, orthogonal time frequency space (OTFS) modulation technology  was proposed  \cite{hadani2017orthogonal,monk2016otfs,8058662} and attracted much attention due to significant advantages in time-variant channels. OTFS converts the signal in the time-frequency domain into a delay-Doppler domain and performs modulation accordingly. In particular, each information symbol in the delay-Doppler domain is expressed by a pair of orthogonal basis functions in the time-frequency domain, i.e., full diversity in the time-frequency domain. 
Furthermore, transmitted symbols experience roughly constant channels even with high Doppler spread. Another advantage is that the implementation of OTFS can be achieved simply as an overlay of the OFDM systems. Farhang \emph{et al}. investigated the OTFS system based on OFDM \cite{8119540} and designed a modem scheme with low complexity, where the cyclic prefix (CP) is added in front of each OFDM symbol in an OTFS symbol.

Channel estimation is crucial in OTFS systems. In \cite{8503182} and \cite{8671740}, training-based channel estimation methods were investigated. In \cite{8503182}, Murali \emph{et al}. used the pseudo-random (PN) sequence as pilots to acquire the channel state information (CSI). In \cite{8671740}, Raviteja \emph{et al}. arranged the pilots and data in the same OTFS symbol and estimated the delay-Doppler channel by a threshold method. 
After the acquisition of the CSI, the symbol detection can be performed. In \cite{8503182}, the Markov chain Monte Carlo (MCMC) sampling  was utilized for the low-complexity symbol detection. In \cite{Raviteja2018Interference}, an explicit derivation of the input-output relationship of OTFS systems was given, and the effect of fractional Doppler caused by insufficient sampling of Doppler-dimension was analyzed. They also developed a message passing (MP) method to detect OTFS symbols. However, when they used the rectangular pulse-shaping waveforms, the CP was not considered, which would cause ISI and might make the receiver more complicated. Different from the symbol spaced sampling framework in \cite{Raviteja2018Interference}, Ge \emph{et al}. designed a fractionally sampling scheme  for symbol detection \cite{FSS}, where only one CP is inserted for the whole OTFS symbol. Compared with \cite{8119540}, less CP overhead in \cite{8671740,Raviteja2018Interference,FSS} can improve spectral efficiency.
In \cite{9321356}, Wei \emph{et al}. reduced the channel spread caused by the fractional Doppler and improved the channel estimation performance through the design of transmitter and receiver windows.
In \cite{8647394}, the embedded pilot-aided CSI acquisition scheme proposed in \cite{8671740} and the MP algorithm proposed in \cite{Raviteja2018Interference} were extended to multiple-input multiple-output (MIMO) OTFS systems.

Massive MIMO technology has been one of the key technologies of 5G owing to the immense improvement of spectrum and power efficiencies \cite{5595728,7913686,6375940,7332961,6415388,7045498,6457363,6736761,9042356}. Considering high Doppler spread scenarios, the integration of massive MIMO and OTFS can further improve the performance.
To utilize such benefits, the base station (BS) requires downlink CSI for precoding.
In traditional time-division duplexing (TDD) systems, uplink training can be used to acquire the downlink CSI via leveraging the channel reciprocity \cite{5595728,6798744}.
However, when systems work in frequency-division duplexing (FDD) mode, user terminals need the pilots transmitted by the BS to acquire the CSI and feed it back. Thus the downlink channel estimation is necessary for massive MIMO-OTFS systems.
In \cite{shen2019channel}, Shen \emph{et al}. demonstrated the 3-dimensional sparsity of the downlink massive MIMO-OTFS channel, based on which a 3D-structured orthogonal matching pursuit (3D-SOMP) method was proposed to acquire the downlink CSI. However, the pilot used for the channel estimation is the random one, which is hard to be implemented and consumes much memory for its randomness in the practical system.
In \cite{9110823}, Liu \emph{et al}. perform the downlink channel estimation with the help of channel parameters obtained from uplink training. But downlink CSI acquisition can not be performed independently and must be after the uplink channel estimation.
In \cite{ShanY}, Shan \emph{et al}. designed a low-overhead pilot pattern for the CSI acquisition and equalized the received signal from each angle through the channel information obtained.  
It is noticed that the fractional Doppler is not considered in \cite{shen2019channel,9110823,ShanY}, and the system models are accurate only when the Doppler frequency of each path of the channel is mapped to an integer tap. It is impractical since the resolution of the Doppler axis is not sufficient.
In \cite{9181410}, Li \emph{et al}. developed a path division multiple access (PDMA) scheme by which different users used scheduled delay-Doppler domain grids to communicate with BS simultaneously without inter-user interference. However, they only considered  TDD systems, where  downlink CSI can be obtained by the reciprocity between the uplink and downlink channel. Thus only the uplink channel estimation was discussed, and the proposed scheme is not suitable for FDD systems.

In this work, we consider the fractional Doppler caused by insufficient sampling of Doppler-dimension in practical systems. Moreover, for massive MIMO-OTFS systems, we propose a downlink CSI acquisition scheme, including deterministic pilot design and channel estimation algorithm. The contributions are summarized as follows:

\begin{itemize}
	\item  In the single-input single-output (SISO) case, we analyze the input-output relationship of the OTFS system, where the OFDM modem is chosen as the time-frequency modem. 
	Specifically, we consider that the Doppler frequency of each path is mapped to an integer and a fractional Dopper tap and reveal the influence of the subpaths contained in each dominant path.
	Moreover, the characteristics of subpaths are used to simplify such a system model.
	Then we extend it to the massive MIMO-OTFS system and establish the downlink CSI acquisition model.
	
	\item To reconstruct the channel accurately, we design a Zadoff-Chu (ZC) sequence based deterministic pilot, which can achieve better sensing performance and memory consumption saving than the random pilot used in \cite{shen2019channel}. 
	We first analyze the relationship between the  pilot matrix and the position of pilots at each beam. Then we give two conditions that the deterministic pilots need to satisfy to ensure the low coherence between the pilot matrix columns. Next, ZC sequences are chosen as pilots, whose performance is also discussed.

	\item Based on the modified model, we propose a modified sensing matrix based channel estimation (MSMCE) algorithm to acquire CSI. Due to the slow change of delay-Doppler channels, path delays and Doppler frequencies of all dominant paths are extracted from the previous channel estimation result and used to modify the sensing matrix for more accurate CSI acquisition. Then we can utilize the estimated CSI to update the path delays and Doppler frequencies, which can be used in the next channel estimation.
	
\end{itemize}

The rest of the paper is organized as follows. 
In Section II, we investigate models for SISO OTFS systems and massive MIMO-OTFS systems. 
Section III establish a downlink CSI acquisition model. Based on the model, we design a deterministic pilot matrix and propose a MSMCE algorithm.
Section IV presents the performance of the proposed CSI acquisition scheme by simulation results, and the paper is concluded in Section V.

\emph{Notations}: 
The superscripts ${\left(  \cdot  \right)^ * }$ and ${\left(  \cdot  \right)^{\rm{H}}}$ denote the conjugate and conjugated-transpose operations, respectively.
The uppercase (lowercase) boldface letters denote matrices (column vectors).
${\bar \jmath} = \sqrt { - 1} $ denotes the imaginary unit.
${\left\| {\bf{X}} \right\|_1}$ and ${\left\| {\bf{x}} \right\|_1}$ denote the $\ell_1$-norm of ${\bf{X}}$ and ${\bf{x}}$, and ${\left\| {\bf{x}} \right\|_2}$ denotes the $\ell_2$-norm of ${\bf{x}}$.
$\lceil x \rceil$ denotes the smallest integer that is not less than $x$, while $\left\lfloor x \right\rfloor $ denotes the largest integer that is not greater than $x$.
$ \odot $ is the Hadamard product operator, and $ \otimes $ is the Kronecker product operator.
${\left(  \cdot  \right)_M}$ denotes mod $M$, and ${\left\langle x \right\rangle _N}$ denotes ${\left( {x + \left\lfloor {\frac{N}{2}} \right\rfloor } \right)_N} - \left\lfloor {\frac{N}{2}} \right\rfloor $.
The notation $ \buildrel \Delta \over = $ is used for definitions, and $\approxeq$ is used to indicate equivalence or approximate equivalence.
$\delta \left(  \cdot  \right)$ denotes the Dirac delta function.
${\left[ {\bf{X}} \right]_{a,:}}$ denotes the $a$-th row of ${\bf{X}}$, wihle ${\left[ {\bf{X}} \right]_{:,b}}$ denotes the $b$-th colomn of ${\bf{X}}$. ${\left[ {\bf{X}} \right]_{a,b}}$ denotes the $(a,b)$-th element of ${\bf{X}}$. ${\left[ {\bf{x}} \right]_a}$ denotes the $a$-th element of ${\bf{x}}$.

\section{System Model}
In this section, some basic definitions and concepts of OTFS are reviewed first. Then we analyze the input-output relationship of the SISO OTFS system, where the OFDM modem is selected as time-frequency modem, and the fractional Doppler is considered. Then we extend it to the case of massive MIMO systems.

\subsection{SISO OTFS System}
For the sake of the subsequent derivation, we first define a lattice in the time-frequency domain as $\Lambda  = \{ nT_{\rm{sym}},m\Delta f\}$,
where $T_{\rm{sym}}$ (seconds) and $\Delta f$ (Hz) are sampling intervals of the time-dimension and the frequency-dimension, respectively, $n = 0, \cdots ,N - 1$, $m = 0, \cdots ,M - 1$, $N,M \in {\mathbb{Z}^ + }$. Similarly, the lattice in the delay-Doppler domain is defined as $\Gamma  = \left\{ {\frac{k}{{NT_{\rm{sym}}}},\frac{l }{{M\Delta f}}} \right\}$
where $\frac{1}{{M\Delta f}}$ and $\frac{1}{{NT_{\rm{sym}}}}$ are sampling intervals of the delay-dimension and Doppler-dimension, respectively, and $k = \left\lceil { - \frac{N}{2}} \right\rceil , \cdots ,\left\lceil {\frac{N}{2}} \right\rceil  - 1$, $l  = 0, \cdots ,M - 1$.

We then arrange a set of $NM$ quadrature amplitude modulated (QAM) symbols ${{X^{{\rm{DD}}}}\left[ {k,l} \right]}$ in the delay-Doppler domain on the lattice $\Gamma$. The inverse symplectic finite Fourier transform (ISFFT) is utilized at OTFS transmitter to convert ${{X^{{\rm{DD}}}}\left[ {k,l} \right]}$ to the symbols ${X^{{\rm{TF}}}}\left[ {n,m} \right]$ in the time-frequency domain  as \cite{1088793}
\begin{equation}\label{equ2-1}
{X^{{\rm{TF}}}}\!\left[ {n,m} \right]\! = \!\frac{1}{{\sqrt {MN} }}\!\sum\limits_{k = \left\lceil { - {N \mathord{\left/
				{\vphantom {N 2}} \right.
				\kern-\nulldelimiterspace} 2}} \right\rceil }^{\left\lceil {{N \mathord{\left/
				{\vphantom {N 2}} \right.
				\kern-\nulldelimiterspace} 2}} \right\rceil  - 1} \!{\sum\limits_{l = 0}^{M - 1} \!{{X^{{\rm{DD}}}}\left[ {k,l} \right]{e^{ - {\bar\jmath}2\pi \left( {\frac{{ml}}{M} - \frac{{nk}}{N}} \right)}}} } .
\end{equation}

Next, an OFDM modulator can be used as a time-frequency modulator to convert ${X^{{\rm{TF}}}}\left[ {n,m} \right]$ to a transmitted signal $s(t)$ with a transmitted waveform ${g_{{\rm{tx}}}}(t)$ as
\begin{align}\label{equ2-2}
s(t)&=\sum\limits_{m = 0}^{M - 1} {\sum\limits_{n = 0}^{N - 1} X^{{\rm{TF}}} } [n,m]{e^{{\bar\jmath}2\pi m\Delta f(t - \frac{{{M_{\rm{CP}}}T}}{M} - n{T_{\rm{sym}}})}} \nonumber \\
& \qquad\qquad\qquad\qquad\qquad\qquad \times{g_{{\rm{tx}}}}(t - n{T_{{\rm{sym}}}}),
\end{align}
where ${{M_{{\rm{CP}}}}}$ is the length of CP, $T = \frac{{M{T_{{\rm{sym}}}}}}{{ {M + {M_{{\rm{CP}}}}} }}$ is time duration of an OFDM symbol without CP, and ${g_{{\rm{tx}}}}(t)$ is defined as 
\begin{equation}\label{equ2-3}
{g_{{\rm{tx}}}}(t) \buildrel \Delta \over = \left\{ {\begin{array}{*{20}{l}}
	{\frac{1}{{\sqrt T }}},&{0 \le t \le {T_{{\rm{sym}}}}}\\
	0,&{{\rm{otherwise}}}
	\end{array}} \right. .
\end{equation}
Note that the CP is added in this step.

After the transmitted signal $s(t)$ passing through the multipath time-variant channel, the received signal $r(t)$ can be obtained as
\begin{equation}\label{equ2-4}
r(t) = \iint {{h_c}(\tau ,\nu ){e^{{\bar\jmath}2\pi \nu (t - \tau )}}s(t - \tau ){\rm{d}}\tau {\rm{d}}\nu  + n(t)} ,
\end{equation}
where $\tau$ is the delay, $\nu$ is the Doppler frequency, ${{h_{\rm{c}}}(\tau ,\nu )}$ is impulse response in the delay-Doppler domain\cite{WCJakes}, and $n(t)$ is the additive Gaussian noise. Usually,  there are only a few scatterers in the transmission environment. Thus the channel can be represented in a sparse way \cite{Raviteja2018Interference}. We assume that the number of the dominant paths between the transmitter and the receiver is $P$, and each consists of $S$ subpaths. Hence, ${{h_{\rm{c}}}(\tau ,\nu )}$ is given by 
\begin{equation}\label{equ2-5}
{h_c}(\tau ,\nu ) = \sum\limits_{i = 0}^{P-1} {\sum\limits_{{s_i} = 0}^{S-1} {{h_{{s_i}}}\delta \left( {\tau  - {\tau _i}} \right)} } \delta \left( {\nu  - {\nu _{{s_i}}}} \right),
\end{equation}
where $h_{{s_i}}$ and ${{\nu _{{s_i}}}}$ are the complex path gain and the Doppler frequency of the $(s_i+1)$-th subpath of the $(i+1)$-th dominant path, respectively. All of subpaths in the $(i+1)$-th dominant path have the same delay ${{\tau _i}}$ \cite{Quadriga}. We define the delay and Doppler taps for $s_i$ subpath as
\begin{equation}\label{equ2-6}
{\tau _i} = \frac{{{l_i}}}{{M\Delta f}},{\nu _{{s_i}}} = \frac{{{k_{{s_i}}} + {{\tilde k}_{{s_i}}}}}{{NT_{\rm{sym}}}},
\end{equation}
where ${{l_i}}$ and ${{k_{{s_i}}}}$ are  integers and represent the indexes of delay and Doppler taps. The real number ${{{\tilde k}_{{s_i}}}}$, whose value range is $\left( { - \frac{1}{2},\frac{1}{2}} \right]$, is defined as the fractional Doppler. The fractional delay is not considered since the resolution $\frac{1}{{M\Delta f}}$ of the delay axis is sufficient so that each path delay can be mapped to an integer delay tap in typical wide-band systems \cite{tse2005fundamentals}.

At the receiver, the cross-ambiguity function between a received waveform ${g_{{\rm{rx}}}}(t)$ and the received signal $r(t)$  is computed by the matched filter as
\begin{equation}\label{equ2-7}
{A_{{g_{{\rm{rx}}}},r}}\left( {\tau ,\nu } \right) = \int {{e^{ - {\bar\jmath}2\pi \nu (t - \tau )}}} g_{{\rm{rx}}}^*(t - \tau )r(t){\rm{d}}t,
\end{equation}
where ${g_{{\rm{rx}}}}(t)$ is defined as 
\begin{equation}\label{equ2-8}
{g_{{\rm{rx}}}}(t) \buildrel \Delta \over = \left\{ {\begin{array}{*{20}{l}}
	{\frac{1}{{\sqrt T }}},&{\frac{{{M_{{\rm{CP}}}}T}}{M} \le t \le {T_{{\rm{sym}}}}}\\
	0,&{{\rm{otherwise}}}
	\end{array}} \right. .
\end{equation}
Note that the CP is removed in this step.
By sampling the cross-ambiguity function, the received data in the time-frequency domain is given by 
\begin{equation}\label{equ2-9}
{Y^{{\rm{TF}}}}\left[ {n,m} \right] = {\left. {{A_{{g_{{\rm{rx}}}},r}}\left( {\tau ,\nu } \right)} \right|_{\tau  = n{T_{{\rm{sym}}}},\nu  = m\Delta f}}.
\end{equation}
It is worth noting that when the time duration of CP, i.e., $\frac{{{M_{{\rm{CP}}}}T}}{M}$ is beyond the maximum path delay of all dominant paths, there is no ISI between OFDM symbols within an OTFS symbol at the receiver, which is different from \cite{8671740,Raviteja2018Interference,FSS}.

Next, the symplectic finite Fourier transform (SFFT) can be utilized to map ${Y^{{\rm{TF}}}}\left[ {n,m} \right]$ to the symbols ${{Y^{{\rm{DD}}}}\left[ {k,l} \right]}$ in the delay-Doppler domain as
\begin{equation}\label{equ2-15}
{Y^{{\rm{DD}}}}\left[ {k,l} \right] = \frac{1}{{\sqrt {NM} }}\sum\limits_{n = 0}^{N - 1} {\sum\limits_{m = 0}^{M - 1} {{Y^{{\rm{TF}}}}\left[ {n,m} \right]{e^{  {\bar\jmath}2\pi \left( {\frac{{ml}}{M} - \frac{{nk}}{N}} \right)}}} }.
\end{equation}
We define  the complex gain of the time-variant channel on the delay tap $l$ at time ${\rho {T_{\rm{s}}}}$ as
\begin{equation}
{h_{\rho ,l }} = \sum\limits_{i = 0}^{P - 1} {\sum\limits_{{s_i} = 0}^{S - 1} {{h_{{s_i}}}{e^{{\bar\jmath}2\pi (\rho-l) {T_{\rm{s}}}{\nu _{{s_i}}}}}\delta \left( {l {T_{\rm{s}}} - {\tau _i}} \right)} } ,
\end{equation}
where ${T_{\rm{s}}} = \frac{1}{{M\Delta f}}$ is the system sampling interval, and we define $h_{\rho ,l_i}^{{s_i}}$ as the complex gain of the $(s_i+1)$-th subpath of ${h_{\rho ,l_i }}$.
Through the above discussion, we give the input-output relationship of the SISO OTFS system, as shown in Proposition 1.

 {\bf Proposition 1}: For a SISO OTFS system, when the OFDM modem is used as the time-frequency modem, the input-output relationship is expressed as
\begin{align}\label{equ2-16}
{Y^{{\rm{DD}}}}\left[ {k,l} \right] &= \frac{1}{\sqrt{N}}\sum\limits_{i = 0}^{P - 1}  \sum\limits_{k' = \left\lceil { - {N \mathord{\left/
				{\vphantom {N 2}} \right.
				\kern-\nulldelimiterspace} 2}} \right\rceil }^{\left\lceil {{N \mathord{\left/
				{\vphantom {N 2}} \right.
				\kern-\nulldelimiterspace} 2}} \right\rceil  - 1}  {H^{{\rm{DD}}}}\left[ {k',{l_i},l} \right] \nonumber \\
			& \qquad \times {X^{{\rm{DD}}}}\left[ {{{\left\langle {k - k'} \right\rangle }_N},{{\left( {l - {l_i}} \right)}_M}} \right] + {Z^{{\rm{DD}}}}\left[ {k,l} \right],
\end{align}
where
\begin{align}\label{equ2-subpath}
&{H^{{\rm{DD}}}}\left[ {k',{l_i},l} \right] \nonumber \\
&\buildrel \Delta \over = \!\frac{1}{{\sqrt N }}\!\!\sum\limits_{j = 0}^{N - 1}\!\! {\left( {\sum\limits_{{s_i} = 0}^{S - 1}\! {h_{{M_{{\rm{CP}}}} + j\left( {M + {M_{{\rm{CP}}}}} \right),{l_i}}^{{s_i}}\!{e^{{\bar\jmath}2\pi \frac{{l\left( {{k_{{s_i}}} + {{\tilde k}_{{s_i}}}} \right)}}{{\left( {M + {M_{{\rm{CP}}}}} \right)N}}}}} } \!\!\right)\!\!{e^{ - {\bar\jmath}2\pi \frac{{k'j}}{N}}}}   ,
\end{align}
and ${Z^{{\rm{DD}}}}\left[ {k,l} \right]$ is the additive noise in the delay-Doppler domain.

\emph{Proof}: See Appendix.

From (\ref{equ2-subpath}), we have the following two findings. First, ${H^{{\rm{DD}}}}\left[ {k',{l_i},l} \right]$ is related to $l$, which is the received data position along the delay-dimension. Second, for the $(i+1)$-th dominant path, ${H^{{\rm{DD}}}}\left[ {k',{l_i},l} \right]$ is affected by all subpaths due to the fractional Doppler. These two points make the system model very complicated, which is not conducive to subsequent analysis. 
Hence, we utilize the characteristics of subpaths to simplify it.
Since all subpaths of one dominant path usually originate from the same scattering cluster, the directions of arrival deviation of these subpaths at the user terminal are usually slight \cite{QuDriGaDoc}. Therefore, we approximate the Doppler frequency of all subpaths of the $(i+1)$-th dominant path to the same value ${\nu _i} = \frac{{{k_i} + {{\tilde k}_i}}}{{N{T_{{\rm{sym}}}}}}$, i.e.,  ${k_{{s_i}}} + {{\tilde k}_{{s_i}}}  \approx {k_i} + {{\tilde k}_i}$, ${s_i} = 0, \cdots ,S - 1$.  Such an approximation simplifies the model in (\ref{equ2-16}) to the follows
\begin{align}\label{equ2-model}
{Y^{{\rm{DD}}}}\left[ {k,l} \right]& \approxeq \frac{1}{\sqrt N}\sum\limits_{i = 0}^{P - 1} {\sum\limits_{k' = \left\lceil {{{ - N} \mathord{\left/
					{\vphantom {{ - N} 2}} \right.
					\kern-\nulldelimiterspace} 2}} \right\rceil }^{\left\lceil {{N \mathord{\left/
					{\vphantom {N 2}} \right.
					\kern-\nulldelimiterspace} 2}} \right\rceil  - 1}\!\!\!\! {{H^{{\rm{DD}}}}\left[ {k',{l_i}} \right]} } {e^{{\bar\jmath}2\pi \frac{{l\left( {{k_i} + {{\tilde k}_i}} \right)}}{{\left( {M + {M_{\rm{CP}}}} \right)N}}}} \nonumber \\
				& \qquad \times {X^{{\rm{DD}}}}\left[ {{{\left\langle {k - k'} \right\rangle }_N},{{\left( {l - {l_i}} \right)}_M}} \right] + {Z^{{\rm{DD}}}}\left[ {k,l} \right],
\end{align}
where ${H^{{\rm{DD}}}}\left[ {k,l} \right]$ is the delay-Doppler domain channel, which is defined as
\begin{equation}\label{equ2-17}
{H^{{\rm{DD}}}}\left[ {k,l} \right] = \frac{1}{\sqrt N}\sum\limits_{j = 0}^{N - 1} {{h_{{M_{\rm{CP}}} + j\left( {M + {M_{\rm{CP}}}} \right),l}}{e^{ - {\bar\jmath}2\pi \frac{{kj}}{N}}}} .
\end{equation}
Equality in (\ref{equ2-model}) holds precisely if there is only one subpath in a dominant path (i.e., $S = 1$).

It is noticed that when $N \to \infty $, (\ref{equ2-16}) is transformed into
\begin{align}\label{equ2-19}
&{Y^{{\rm{DD}}}}\!\left[ {k,l} \right]\mathop  = \limits^{N \to \infty } \!\frac{1}{\sqrt N}\!\sum\limits_{l' = 0}^{M - 1} {\sum\limits_{k' = \left\lceil {{{ - N} \mathord{\left/
					{\vphantom {{ - N} 2}} \right.
					\kern-\nulldelimiterspace} 2}} \right\rceil }^{\left\lceil {{N \mathord{\left/
					{\vphantom {N 2}} \right.
					\kern-\nulldelimiterspace} 2}} \right\rceil  - 1} \!\!\!\!{{H^{{\rm{DD}}}}\left[ {k',l'} \right]} } {e^{{\bar\jmath}2\pi \frac{{lk'}}{{\left( {M + {M_{\rm{CP}}}} \right)N}}}} \nonumber \\
&\qquad\qquad\qquad\quad\times{X^{{\rm{DD}}}}\!\left[ {{{\left\langle {k - k'} \right\rangle }_N},{{\left( {l - l'} \right)}_M}} \right] \!+\! {Z^{{\rm{DD}}}}\left[ {k,l} \right],
\end{align}
which is the same as that in \cite{shen2019channel} and \cite{9110823}. Therefore, the model in \cite{shen2019channel} and \cite{9110823} can be seen as a special case  of our proposed input-output relationship when $N \to \infty $. Moreover, comparing (\ref{equ2-model}) with (\ref{equ2-19}), we can find that the phase compensation ${e^{{\bar\jmath}2\pi \frac{{l\left( {{k_i} + {{\tilde k}_i}} \right)}}{{\left( {M + {M_{\rm{CP}}}} \right)N}}}}$ in (\ref{equ2-model}) uses the information of each dominant path, which makes the model more accurate in the practical system.

Note that we choose the OFDM-based OTFS system similar to \cite{8119540} in this paper, which can be achieved simply as an overlay of the widely used OFDM system, instead of the spectral efficient OTFS model in \cite{8671740,Raviteja2018Interference,FSS}. If fewer CPs are used to improve spectral efficiency as in \cite{8671740,Raviteja2018Interference,FSS}, the input-output relationship will have an additional case owing to the ISI between OFDM symbols within an OTFS symbol, and the only difference between the two cases is the exponential term, which implies that the basic idea of using the channel path information to modify the phase compensation matrix of the sensing matrix to improve the channel estimation performance, as shown later, is still valid in the spectral efficient OTFS model.

\subsection{Massive MIMO-OTFS System}
We consider the downlink transmission in a massive MIMO-OTFS system.
The BS deploys $N_{\rm{t}}$ antennas and serves $U$ single-antenna user terminals. We consider the centralized massive MIMO, where the uniform linear array (ULA) is equipped at the BS, and the antenna spacing is set to half wavelength.
Without loss of generality, we focus on one user terminal and disregard the dependency of the channel on user index for simplicity. 

Similar to (\ref{equ2-model}), the symbols received at the user terminal in the delay-Doppler domain can be expressed as
\begin{align}\label{equ2-model1}
&{Y^{{\rm{DD}}}}\left[ {k,l} \right] \nonumber\\
&\approxeq \frac{1}{{\sqrt N }}\sum\limits_{{n_{\rm{t}}} = 0}^{{N_{\rm{t}}} - 1} \sum\limits_{i = 0}^{P - 1}  \sum\limits_{k' = \left\lceil { - {N \mathord{\left/
					{\vphantom {N 2}} \right.
					\kern-\nulldelimiterspace} 2}} \right\rceil }^{\left\lceil {{N \mathord{\left/
					{\vphantom {N 2}} \right.
					\kern-\nulldelimiterspace} 2}} \right\rceil  - 1} \!\! {H^{{\rm{DDS}}}}\left[ {k',{l_i},{n_{\rm{t}}}} \right]{e^{{\bar\jmath}2\pi \frac{{l\left( {{k_i} + {{\tilde k}_i}} \right)}}{{\left( {M + {M_{{\rm{CP}}}}} \right)N}}}} \nonumber \\ 
				&\quad\quad\times {X^{{\rm{DDS}}}}\left[ {{{\left\langle {k - k'} \right\rangle }_N},{{\left( {l - {l_i}} \right)}_M},{n_{\rm{t}}}} \right]  + {Z^{{\rm{DD}}}}\left[ {k,l} \right] ,
\end{align}
where ${X^{{\rm{DDS}}}}\left[ {k,l,{n_{\rm{t}}}} \right]$ is  transmitted symbols in the delay-Doppler-space domain, ${H^{{\rm{DDS}}}}\left[ {k',{l_i},{n_{\rm{t}}}} \right]$ is the delay-Doppler-space domain channel, which is defined as \cite{hlawatsch2011wireless}
\begin{align}
&{H^{{\rm{DDS}}}}\left[ {k,l,{n_{\rm{t}}}} \right] \nonumber \\
&\quad= \frac{1}{\sqrt{N}}\sum\limits_{j = 0}^{N - 1}  \sum\limits_{i = 0}^{P - 1} \sum\limits_{{s_i} = 0}^{S - 1} {h_{{s_i}}}{e^{{\bar \jmath}2\pi \left( {{M_{{\rm{CP}}}} + j\left( {M + {M_{{\rm{CP}}}}} \right)}-l \right){T_{\rm{s}}}{\nu _{{s_i}}}}} \nonumber \\
			&\qquad\qquad\qquad\times \delta \left( {\ell {T_{\rm{s}}} - {\tau _i}} \right){e^{{\bar \jmath}\pi {n_{\rm{t}}}\sin {\varphi _{{s_i}}}}}  {e^{ - {\bar \jmath}2\pi \frac{{kj}}{N}}} ,
\end{align}
where ${\varphi _{{s_i}}} \in \left[ { - {\pi  \mathord{\left/{\vphantom {\pi  {2,{\pi  \mathord{\left/{\vphantom {\pi  2}} \right.\kern-\nulldelimiterspace} 2}}}} \right.\kern-\nulldelimiterspace} {2,{\pi  \mathord{\left/{\vphantom {\pi  2}} \right.\kern-\nulldelimiterspace} 2}}}} \right)$ is the angle of departure (AoD) of the $(s_i+1)$-th subpath. To exploit the sparsity of the beam domain, the delay-Doppler-beam domain channel $H^{\rm{DDB}}\left[ {k,l,b} \right]$ is obtained by applying normalized discrete Fourier transform (DFT) for the delay-Doppler-space domain channel along the space-dimension $n_{\rm{t}}$ as \cite{shen2019channel,sun2015beam}
\begin{align}\label{equ2-22}
&{H^{{\rm{DDB}}}}\left[ {k,l,b} \right] \nonumber \\
&\ = \frac{1}{{\sqrt {{N_{\rm{t}}}} }}\sum\limits_{{n_{\rm{t}}} = 0}^{{N_{\rm{t}}} - 1} {{H^{{\rm{DDS}}}}\left[ {k,l,{n_{\rm{t}}}} \right]} {e^{ - {\bar\jmath}2\pi \frac{{b{n_{\rm{t}}}}}{{{N_{\rm{t}}}}}}}\nonumber\\
&\ = \!\sqrt {\frac{N}{{{N_{\rm{t}}}}}} \sum\limits_{i = 0}^{P - 1} {\sum\limits_{{s_i} = 0}^{S - 1} {{h_{{s_i}}}{e^{{\bar\jmath}2\pi ({M_{{\rm{CP}}}}-l){T_{\rm{s}}}{\nu _{{s_i}}}}}{\Xi _N}\!\left( {k - N{T_{{\rm{sym}}}}{\nu _{{s_i}}}} \right)} } \nonumber \\
&\qquad\qquad\qquad\times\delta \left( {l {T_{\rm{s}}} - {\tau _i}} \right){\Xi _{N_{\rm{t}}}}\left( {b - {{{N_{\rm{t}}}\sin {\varphi _{{s_i}}}} \mathord{\left/
			{\vphantom {{{N_{\rm{t}}}\sin {\varphi _{{s_i}}}} 2}} \right.
			\kern-\nulldelimiterspace} 2}} \right),
\end{align}
where $b =  - \frac{{{N_{\rm{t}}}}}{2}, \cdots ,0, \cdots ,\frac{{{N_{\rm{t}}}}}{2} - 1$ is the beam index, and ${\Xi _N}(x) \buildrel \Delta \over = \frac{1}{N}\sum\limits_{i = 0}^{N - 1} {{e^{ - {\bar\jmath}2\pi \frac{x}{N}i}}} $. 
From (\ref{equ2-22}), we can find that the dominant elements of ${H^{{\rm{DDB}}}}\left[ {k,l,b} \right]$ are distributed in the positions where $k \approx N{T_{{\rm{sym}}}}{\nu _{{s_i}}}$, $l \approx {\tau _i}M\Delta f$ and $b \approx {{{N_{\rm{t}}}\sin {\varphi _{{s_i}}}} \mathord{\left/{\vphantom {{{N_{\rm{t}}}\sin {\varphi _{{s_i}}}} 2}} \right.
		\kern-\nulldelimiterspace} 2}$, which means that the channel has the 3D sparsity over the delay-Doppler-beam domain \cite{shen2019channel,5454399}. 
By combining (\ref{equ2-model1}) and (\ref{equ2-22}), the received symbols can be rewritten as
\begin{align}\label{equ2-model2}
&{Y^{{\rm{DD}}}}\left[ {k,l} \right] \nonumber \\
& \approxeq \frac{1}{{\sqrt N }}\!\sum\limits_{b =  - {{{N_{\rm{t}}}} \mathord{\left/
			{\vphantom {{{N_{\rm{t}}}} 2}} \right.
			\kern-\nulldelimiterspace} 2}}^{{{{N_{\rm{t}}}} \mathord{\left/
			{\vphantom {{{N_{\rm{t}}}} 2}} \right.
			\kern-\nulldelimiterspace} 2} - 1} {\sum\limits_{i = 0}^{P - 1}  \sum\limits_{k' = \left\lceil { - {N \mathord{\left/
					{\vphantom {N 2}} \right.
					\kern-\nulldelimiterspace} 2}} \right\rceil }^{\left\lceil {{N \mathord{\left/
					{\vphantom {N 2}} \right.
					\kern-\nulldelimiterspace} 2}} \right\rceil  - 1} \!\!\! {H^{{\rm{DDB}}}}\left[ {k',{l_i},b} \right]{e^{{\bar\jmath}2\pi \frac{{l\left( {{k_i} + {{\tilde k}_i}} \right)}}{{\left( {M + {M_{{\rm{CP}}}}} \right)N}}}}} \nonumber \\ &\quad\quad\times{X^{{\rm{DDB}}}}\left[ {{{\left\langle {k - k'} \right\rangle }_N},{{\left( {l - {l_i}} \right)}_M},b} \right] + {Z^{{\rm{DD}}}}\left[ {k,l} \right],
\end{align}
where 
\begin{equation}\label{equ2-24}
{ X^{{\rm{DDB}}}}\left[ {k,l,b} \right] \buildrel \Delta \over = \frac{1}{{\sqrt {{N_{\rm{t}}}} }}\sum\limits_{{n_{\rm{t}}} = 0}^{{N_{\rm{t}}}} {{e^{{\bar\jmath}2\pi \frac{{r{n_{\rm{t}}}}}{{{N_{\rm{t}}}}}}}} {X^{{\rm{DDS}}}}\left[ {k,l,{n_{\rm{t}}}} \right].
\end{equation}
Equality in (\ref{equ2-model2}) holds precisely if there is only one subpath in a dominant path.
Similarly, when $N \to \infty $, the received symbols are given by
\begin{align}\label{equ2-25}
&{Y^{{\rm{DD}}}}\left[ {k,l} \right] \nonumber \\
&\mathop  = \limits^{N \to \infty } \!\!\frac{1}{\sqrt N}\!\! \sum\limits_{b =  - {{{N_{\rm{t}}}} \mathord{\left/
			{\vphantom {{{N_{\rm{t}}}} 2}} \right.
			\kern-\nulldelimiterspace} 2}}^{{{{N_{\rm{t}}}} \mathord{\left/
			{\vphantom {{{N_{\rm{t}}}} 2}} \right.
			\kern-\nulldelimiterspace} 2} - 1}\! {\sum\limits_{l' = 0}^{M - 1}\! {\sum\limits_{k' = \left\lceil {{{ - N} \mathord{\left/
						{\vphantom {{ - N} 2}} \right.
						\kern-\nulldelimiterspace} 2}} \right\rceil }^{\left\lceil {{N \mathord{\left/
						{\vphantom {N 2}} \right.
						\kern-\nulldelimiterspace} 2}} \right\rceil  - 1}\!\!\!\!\!{H^{{\rm{DDB}}}}\!\left[ {k',l',b} \right]\!{{e^{{\bar\jmath}2\pi \frac{{lk'}}{{\left( {M + {M_{{\rm{CP}}}}} \right)N}}}}} } } \nonumber\\
					&\qquad\quad \times{ X^{{\rm{DDB}}}}\left[ {{{\left\langle {k - k'} \right\rangle }_N},{{\left( {l - l'} \right)}_M},b} \right] + {Z^{{\rm{DD}}}}\left[ {k,l} \right].
\end{align}

According to downlink massive MIMO-OTFS system models and the
sparsity of the delay-Doppler-beam domain channel ${H^{{\rm{DDB}}}}\left[ {k,l,b} \right]$, the downlink CSI acquisition is actually a sparse signal reconstruction problem \cite{shen2019channel}. A variety of compressive sensing (CS) algorithms can be used to solve this problem.
Next, we will give the model of downlink CSI acquisition and propose the corresponding deterministic pilot design and MSMCE algorithm.

\section{Downlink CSI acquisition for Massive MIMO-OTFS Systems}
In this section, we first model the downlink CSI acquisition as a sparse signal reconstruction problem. To improve sensing performance, the deterministic pilots are designed, and the pilot overhead is also discussed.
Then, we propose a MSMCE algorithm to reconstruct the sparse delay-Doppler-beam domain channel and analyze its performance.

\subsection{Downlink CSI Acquisition Model}

\begin{figure}[htbp]
	\centering
	\includegraphics[width=0.5\textwidth]{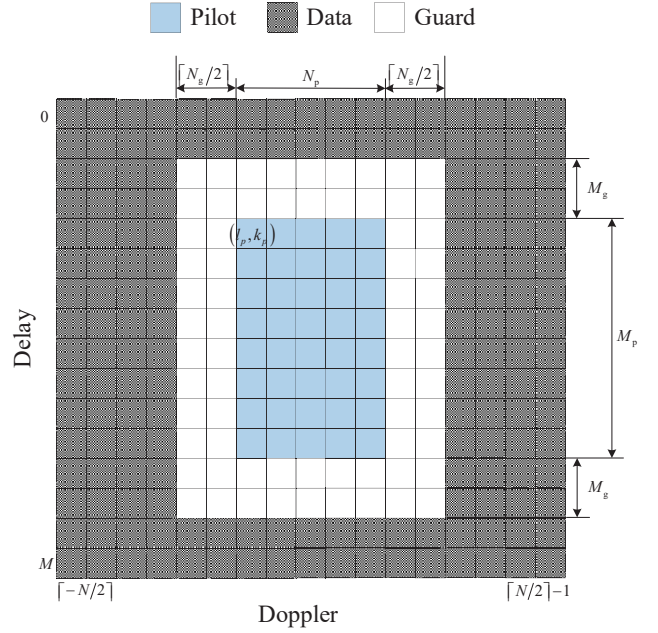}
	\caption{The position of pilots, data, and guard intervals in the delay-Doppler domain at one antenna.}
	\label{fig-ini}
\end{figure}

Fig. \ref{fig-ini} shows the position of pilots in the delay-Doppler domain at one antenna. 
We consider that the position of pilots in the delay-Doppler domain at each transmit antenna is the same and given by
\begin{equation}\label{equ3-1}
k = {k_{\rm{p}}}, \cdots ,{k_{\rm{p}}} + {N_{\rm{p}}} - 1,{\rm{  }}l = {l_{\rm{p}}}, \cdots ,{l_{\rm{p}}} + {M_{\rm{p}}} - 1,
\end{equation}
where ${k_{\rm{p}}}$ and ${l_{\rm{p}}}$ are the initial position of the pilots in the Doppler domain and delay domain, respectively, and ${N_{\rm{p}}}$ and ${M_{\rm{p}}}$ are the lengths of pilots. The guard intervals (i.e., zero symbols) are demanded to eliminate the interference between the data and pilots. 

Delay-Doppler-beam domain channels have finite support $\left[ {0,{\tau _{\max }}} \right]$ and $\left[ { - {\nu _{\max }},{\nu _{\max }}} \right]$ along the delay-dimension and the Doppler-dimension, respectively \cite{hadani2017orthogonal}. Hence for the delay-dimension, the guard intervals should be placed at the beginning and the end of the pilots with the length ${M_{\rm{g}}} \ge {\tau _{\max }}M\Delta f$. 
However, for the Doppler-dimension, ${\nu _{\max }}N{T_{{\rm{sym}}}}$ is typically non-integer in practical systems, which means that according to the characteristic of  ${\Xi _N}(x)$ in (\ref{equ2-22}), the magnitude at each $k$ is not zero and decreases as $k$ moves away from ${\nu _{\max }}N{T_{{\rm{sym}}}}$. Thus we only consider $k \in \left[ { - \frac{{{N_{\max }}}}{2},\frac{{{N_{\max }}}}{2} - 1} \right]$ for ${H^{{\rm{DDB}}}}$ and replace the values outside the range with zero, and the guard intervals should be placed at the beginning and the end of the pilots with the length $\frac{{{N_{\rm{g}}}}}{2} \ge \frac{{{N_{\max }}}}{2}$.
Therefore, the range of $l'$ and $k'$ of the delay-Doppler-beam domain channel ${H^{{\rm{DDB}}}}\left[ {k',l',b} \right]$ are limited to $\left[ {0,{M_{\rm{g}}}} \right]$ and $\left[ {\left\lceil { - {{{N_{\rm{g}}}} \mathord{\left/
				{\vphantom {{{N_{\rm{g}}}} 2}} \right.
				\kern-\nulldelimiterspace} 2}} \right\rceil ,\left\lceil {{{{N_{\rm{g}}}} \mathord{\left/
				{\vphantom {{{N_{\rm{g}}}} 2}} \right.
				\kern-\nulldelimiterspace} 2}} \right\rceil  - 1} \right]$, respectively. Moreover, the data is placed in the position except for the pilot and the guard interval.

The position of received symbols for channel estimation is the same as (\ref{equ3-1}).
Therefore, (\ref{equ2-model2}) can be rewritten as 
\begin{equation}\label{equ3-2}
{{\bf{y}}^{{\rm{DD}}}} \approxeq \left(\bf{\Phi}  \odot {{\bf{X}}^{{\rm{DDB}}}}\right){{\bf{h}}^{{\rm{DDB}}}} + {{\bf{z}}^{{\rm{DD}}}},
\end{equation}
where ${{\bf{X}}^{{\rm{DDB}}}} \in {\mathbb{C}^{{M_{\rm{p}}}{N_{\rm{p}}} \times {M_{\rm{g}}}{N_{\rm{g}}}{N_{\rm{t}}}}}$ is the pilot matrix with element ${X^{{\rm{DDB}}}}\left[ {{{\left\langle {k - k'} \right\rangle }_N},{{\left( {l - l'} \right)}_M},b} \right]$ of index  $({\left( {l - {l_{\rm{p}}}} \right){N_{\rm{p}}} + \left( {k - {k_{\rm{p}}}} \right)  + 1},{(b+{N_{\rm{t}}}/2){M_{\rm{g}}}{N_{\rm{g}}}}+
l'{N_{\rm{g}}} + k' + \left\lfloor {{{{N_{\rm{g}}}} \mathord{\left/
			{\vphantom {{{N_{\rm{p}}}} 2}} \right.
			\kern-\nulldelimiterspace} 2}} \right\rfloor  + 1)$, where $k = {k_{\rm{p}}}, \cdots ,{k_{\rm{p}}} + {N_{\rm{p}}} - 1$, ${\rm{ }}l = {l_{\rm{p}}}, \cdots ,{l_{\rm{p}}} + {M_{\rm{p}}} - 1$, $b =  - \frac{{{N_{\rm{t}}}}}{2}, \cdots ,0, \cdots ,\frac{{{N_{\rm{t}}}}}{2} - 1$, $k' = \left\lceil { - \frac{{{N_{\rm{g}}}}}{2}} \right\rceil , \cdots ,0, \cdots ,$ $\left\lceil {\frac{{{N_{\rm{g}}}}}{2}} \right\rceil  - 1$  and ${\rm{ }}l' = 0, \cdots ,{M_{\rm{g}}} - 1$.		
 In (\ref{equ3-2}), ${{\bf{y}}^{{\rm{DD}}}} \in {\mathbb{C}^{{M_{\rm{p}}}{N_{\rm{p}}} \times 1}}$ and ${{\bf{z}}^{{\rm{DD}}}} \in {\mathbb{C}^{{M_{\rm{p}}}{N_{\rm{p}}} \times 1}}$ are the vector forms of received symbols and the additive noise with element ${Y^{{\rm{DD}}}}\left[ {k,l} \right]$ and ${Z^{{\rm{DD}}}}\left[ {k,l} \right]$ of index $( \left( {l - {l_{\rm{p}}}} \right){N_{\rm{p}}} + \left( {k - {k_{\rm{p}}}} \right)   + 1 )$. ${{\bf{h}}^{{\rm{DDB}}}} \in {\mathbb{C}^{{M_{\rm{g}}}{N_{\rm{g}}}{N_{\rm{t}}} \times 1}}$ is the vector form of the delay-Doppler-beam domain channel with element ${H^{{\rm{DDB}}}}\left[ {k',l',b} \right]$ of index 
$( (b+{N_{\rm{t}}}/2){M_{\rm{g}}}{N_{\rm{g}}} + l'{N_{\rm{g}}} + k' + \left\lfloor {{{{N_{\rm{g}}}} \mathord{\left/
				{\vphantom {{{N_{\rm{g}}}} 2}} \right.
				\kern-\nulldelimiterspace} 2}} \right\rfloor  + 1)$,
${\bf{\Phi }}  \in {\mathbb{C}^{{M_{\rm{p}}}{N_{\rm{p}}} \times {M_{\rm{g}}}{N_{\rm{g}}}{N_{\rm{t}}}}}$ is the phase compensation matrix with element $\phi \left( {l,l'} \right)$ of index $({\left( {l - {l_{\rm{p}}}} \right){N_{\rm{p}}} + \left( {k - {k_{\rm{p}}}} \right)}$ ${ + \left\lfloor {{{{N_{\rm{p}}}} \mathord{\left/
				{\vphantom {{{N_{\rm{p}}}} 2}} \right.
				\kern-\nulldelimiterspace} 2}} \right\rfloor  + 1,(b+{N_{\rm{t}}}/2){M_{\rm{g}}}{N_{\rm{g}}} + l'{N_{\rm{g}}} + k'\! +\! \left\lfloor {{{{N_{\rm{g}}}} \mathord{\left/
				{\vphantom {{{N_{\rm{p}}}} 2}} \right.
				\kern-\nulldelimiterspace} 2}} \right\rfloor  \!+ \!1})$, which is defined as
\begin{equation}\label{equ3-3}
\phi \left( {l,l'} \right) = \left\{ {\begin{array}{*{20}{l}}
	{{e^{{\bar\jmath}2\pi \frac{{l\left( {{k_i} + {{\tilde k}_i}} \right)}}{{\left( {M + {M_{{\rm{CP}}}}} \right)N}}}}},&{l' \in \mbox{Del},l' = {l_i}}\\
	1,&{l' \notin\mbox{Del}}
	\end{array}} \right.,
\end{equation}
where $\mbox{Del}$ is the tap index set of the delay of all dominant paths, i.e., $\mbox{Del} = \left\{ {{l_0},{l_1}, \cdots ,{l_{P - 1}}} \right\}$. ${{k_i} + {{\tilde k}_i}}$ is selected from the set $\mbox{Dop}$ which consists of the integer and fractional tap indexes of the Doppler frequency of all dominant paths, i.e., $\mbox{Dop} = \left\{ {{{k_0} + {{\tilde k}_0}},{{k_1} + {{\tilde k}_1}}, \cdots ,{{k_{P-1}} + {{\tilde k}_{P-1}}}} \right\}$.
When $N \to \infty $, according to (\ref{equ2-25}), ${\bf{\Phi }}$ in (\ref{equ3-2}) is converted to  ${\bf{\tilde \Phi }}  \in {\mathbb{C}^{{M_{\rm{p}}}{N_{\rm{p}}} \times {M_{\rm{g}}}{N_{\rm{g}}}{N_{\rm{t}}}}}$ with element $\tilde \phi \left( {l,k'} \right) = {e^{{\bar\jmath}2\pi \frac{{lk'}}{{\left( {M + {M_{{\rm{CP}}}}} \right)N}}}}$ of index $({\left( {l - {l_{\rm{p}}}} \right){N_{\rm{p}}} + \left( {k - {k_{\rm{p}}}} \right)}$ $   + 1,(b+{N_{\rm{t}}}/2){M_{\rm{g}}}{N_{\rm{g}}} + l'{N_{\rm{g}}} + k'\! +\! \left\lfloor {{{{N_{\rm{g}}}} \mathord{\left/
				{\vphantom {{{N_{\rm{p}}}} 2}} \right.
				\kern-\nulldelimiterspace} 2}} \right\rfloor  \!+ \!1)$.
And in this case, the model is the same as that in \cite{shen2019channel}.

We can rewrite (\ref{equ3-2}) as
\begin{equation}\label{equ3-5}
{{\bf{y}}^{{\rm{DD}}}} \approxeq {\bf{\Theta }}{{\bf{h}}^{{\rm{DDB}}}} + {{\bf{z}}^{{\rm{DD}}}},
\end{equation}
where ${\bf{\Theta }} \buildrel \Delta \over = {\bf{\Phi }} \odot {{\bf{X}}^{{\rm{DDB}}}}$. Thus the downlink CSI acquisition of  massive MIMO-OTFS systems is converted to a sparse signal reconstruction problem, where ${{\bf{h}}^{{\rm{DDB}}}}$ is the sparse vector to be recovered, and ${\bf{\Theta }}$ is the sensing matrix. When $N \to \infty $, ${\bf{\Theta }}$ is converted to ${\bf{\tilde \Theta }} \buildrel \Delta \over = {\bf{\tilde \Phi }} \odot {{\bf{X}}^{{\rm{DDB}}}}$.
Hence we have two kinds of sensing matrix that can be used in CS, and the only difference between them is the phase compensation matrix. The phase compensation matrix ${\bf{ \Phi }}$ is more accurate but requires both path delay and Doppler frequency of each dominant path, which can not be directly obtained. In contrast, the phase compensation matrix  ${\bf{\tilde \Phi }}$ is irrelevant to the channel, although it is inaccurate. To combine the advantages of both, the MSMCE algorithm is proposed and will be discussed in detail later. Prior to that, we give the deterministic pilot design in the next subsection.

\subsection{Deterministic Pilot Design}
In order to recover the sparse vector reliably, the sensing matrix must be designed carefully and satisfies the restricted isometry property (RIP) proposed in \cite{1542412}. The RIP implies that the coherence (i.e., inner product) between  columns in a sensing matrix  should be as small as possible to obtain a good sensing performance.
Although the random matrix is usually used as the sensing matrix due to its near-optimality\cite{4472240}, it is hard to be implemented and costs a lot of memory consumption for its randomness. Therefore, we focus on the deterministic pilot design for practical communication systems.

Since the phase compensation matrix $\bf{\Phi}$ of the sensing matrix ${\bf{\Theta }}$ is related to the channel, which varies with time, the pilot matrix ${{\bf{X}}^{{\rm{DDB}}}}$ is what we analyzed exactly. The design of pilot matrix ${{\bf{X}}^{{\rm{DDB}}}}$ is equivalent to the design of transmitted pilots ${X^{{\rm{DDB}}}}\left[ {k,l,b} \right]$ in the delay-Doppler-beam domain with their position in the delay-Doppler domain at each beam. Note that the pilots ${X^{{\rm{DDS}}}}\left[ {k,l,{n_{\rm{t}}}} \right]$ in the delay-Doppler-space domain  (i.e., the pilots in the delay-Doppler domain at each BS transmit antenna) can be obtained using (\ref{equ2-24}), and such a transformation will not affect the position of  pilots in the delay-Doppler domain.

\begin{figure}[htbp]
	\centering
	\subfigure[]{
		\begin{minipage}{7.0cm}
			\centering
			\includegraphics[width=1\textwidth]{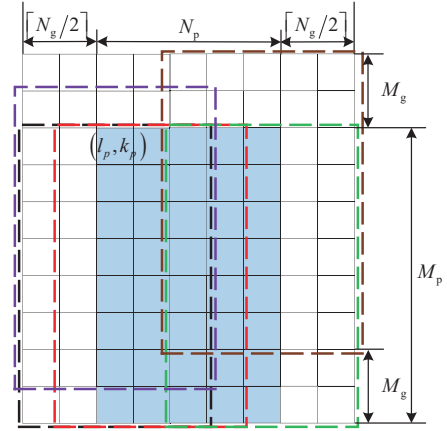}
			\label{fig-con1}
		\end{minipage}
	
	}
	\subfigure[]{
		\begin{minipage}{7.0cm}
			\centering
			\includegraphics[width=1\linewidth]{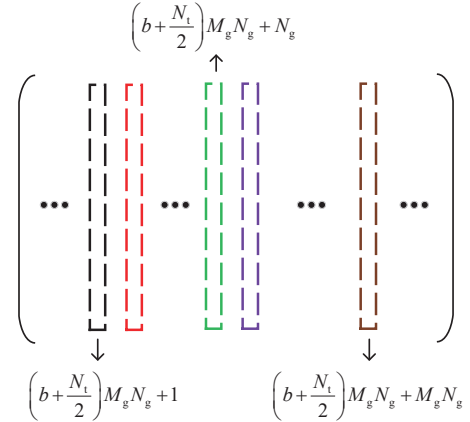}
			\label{fig-con2}
		\end{minipage}
	}
	\caption{The relationship between the position of pilots at one beam and the pilot matrix. (a) Position of pilots in the delay-Doppler domain at the $\left( {b + \frac{{{N_{\rm{t}}}}}{2}} + 1 \right)$-th beam; (b) Pilot matrix.}
	\label{fig-con}
\end{figure}

We first analyze the relationship between the  pilot matrix ${{\bf{X}}^{{\rm{DDB}}}}$ and the pilot position at one beam. As shown in Fig. \ref{fig-con1}, each dashed box contains ${M_{\rm{p}}}{N_{\rm{p}}}$ pilot symbols, which are used to construct the corresponding column of the pilot matrix in Fig. \ref{fig-con2}. The pilots at each beam can construct ${M_{\rm{g}}}{N_{\rm{g}}}$ columns totally.

\begin{figure}
	\centering
	\includegraphics[width=0.5\textwidth]{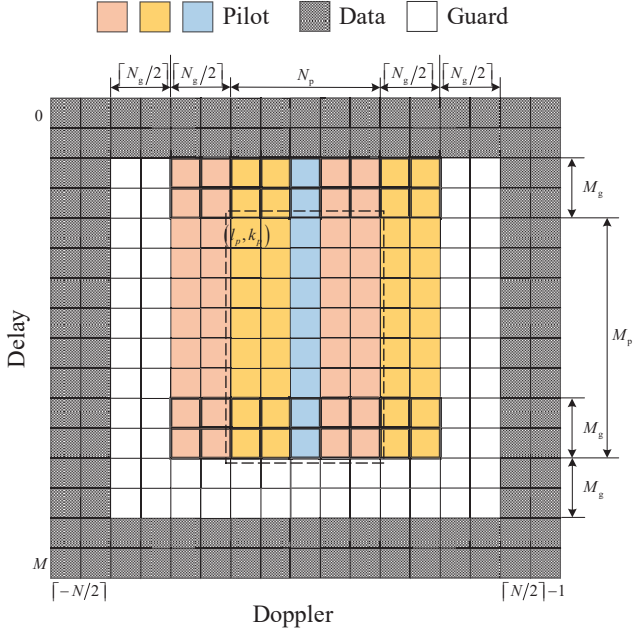}
	\caption{The designed position of pilots, data, and guard intervals in the delay-Doppler domain at one beam.}
	\label{fig-new}
\end{figure}

However, we can find that if we directly use the position of pilots  shown in Fig. \ref{fig-ini}, there will be some zero symbols in the pilot matrix, which will affect the coherence between columns of the pilot matrix and make the pilot design difficult. Therefore, based on Fig. \ref{fig-ini}, we propose a new pilot position design shown in Fig. \ref{fig-new}.
Specifically, along the Doppler-dimension, the first $\left\lceil {{{{N_{\rm{g}}}} \mathord{\left/
			{\vphantom {{{N_{\rm{g}}}} 2}} \right.
			\kern-\nulldelimiterspace} 2}} \right\rceil $ pilots 
are added to the end (marked in yellow grids in Fig. \ref{fig-new}), and the last $\left\lceil {{{{N_{\rm{g}}}} \mathord{\left/{\vphantom {{{N_{\rm{g}}}} 2}} \right.\kern-\nulldelimiterspace} 2}} \right\rceil $ pilots are added to the beginning (marked in pink grids in Fig. \ref{fig-new}) while the guard intervals are placed at the beginning and the end of pilots with the length $\left\lceil {{{{N_{\rm{g}}}} \mathord{\left/{\vphantom {{{N_{\rm{g}}}} 2}} \right.\kern-\nulldelimiterspace} 2}} \right\rceil $. Along the delay-dimension, the last ${{M_{\rm{g}}}}$ pilots are added to the beginning (marked in thick-black grids in Fig. \ref{fig-new}) while the guard intervals are placed at the end of the pilots with the length of ${{M_{\rm{g}}}}$. The data is placed in the position except for the pilot  and the guard interval.

Next, we discuss the pilot design at a given beam and between different beams.  For a given beam, we assume that the $( {\left( {b + \frac{{{N_{\rm{t}}}}}{2}} \right){M_{\rm{g}}}{N_{\rm{g}}} + 1} )$-th column of the pilot matrix (i.e., the black dashed box in Fig. \ref{fig-con2})  constructed by the pilots at this beam shown in Fig. \ref{fig-new} is defined as
\begin{equation}
{\rm{black}}: \ {{\bf{c}}_{\left( {b + \frac{{{N_{\rm{t}}}}}{2}} \right){M_{\rm{g}}}{N_{\rm{g}}} + 1}} = {\bf{p}}_0^{{M_{\rm{p}}}} \otimes {\bf{p}}_0^{{N_{\rm{p}}}},
\end{equation}
where ${\bf{p}}_c^L$ denotes a sequence of length $L$ and is cyclically shifted by $c$ symbols. According to the designed pilot position and its relationship with the pilot matrix, different columns (i.e., dashed boxes with different colors in Fig. \ref{fig-con2}) can be expressed as
\begin{align}
\begin{array}{*{20}{l}}
&{{\rm{red}}:\quad \ {{\bf{c}}_{\left( {b + \frac{{{N_{\rm{t}}}}}{2}} \right){M_{\rm{g}}}{N_{\rm{g}}} + 2}}\ \ \ \ \,  = {\bf{p}}_0^{{M_{\rm{p}}}} \otimes {\bf{p}}_1^{{N_{\rm{p}}}}} \\
&{{\rm{green}}:\ {{\bf{c}}_{\left( {b + \frac{{{N_{\rm{t}}}}}{2}} \right){M_{\rm{g}}}{N_{\rm{g}}} + {N_{\rm{g}}}}}\ \ \  = {\bf{p}}_0^{{M_{\rm{p}}}} \otimes {\bf{p}}_{{N_{\rm{g}}} - 1}^{{N_{\rm{p}}}}}\\
&{{\rm{purple:}}\  \!{{\bf{c}}_{\left( {b + \frac{{{N_{\rm{t}}}}}{2}} \right){M_{\rm{g}}}{N_{\rm{g}}} + {N_{\rm{g}}} + 1}} = {\bf{p}}_1^{{M_{\rm{p}}}} \otimes {\bf{p}}_0^{{N_{\rm{p}}}}}\\
&{{\rm{brown}}:{{\bf{c}}_{\left( {b + \frac{{{N_{\rm{t}}}}}{2}} \right){M_{\rm{g}}}{N_{\rm{g}}} + {M_{\rm{g}}}{N_{\rm{g}}}}} = {\bf{p}}_{{M_{\rm{g}}} - 1}^{{M_{\rm{p}}}} \otimes {\bf{p}}_{{N_{\rm{g}}} - 1}^{{N_{\rm{p}}}}}
\end{array}.
\end{align}
Therefore, according to the property of Kronecker product $\left( {{\bf{A}} \otimes {\bf{B}}} \right)\left( {{\bf{C}} \otimes {\bf{D}}} \right) = \left( {{\bf{AC}}} \right) \otimes \left( {{\bf{BD}}} \right)$, 
the pilot sequence should be orthogonal to its cyclically shifted version to ensure the orthogonality between columns of the pilot matrix at the given beam, which is defined as the \emph{orthogonality condition}.

The above pilot design can be used at up to ${\eta _{{\rm{del}}}}{\eta _{{\rm{dop}}}}$ beams, where ${\eta _{{\rm{del}}}} = \left\lfloor {{{{M_{\rm{p}}}} \mathord{\left/
			{\vphantom {{{M_{\rm{p}}}} {{M_{\rm{g}}}}}} \right.
			\kern-\nulldelimiterspace} {{M_{\rm{g}}}}}} \right\rfloor $ and ${\eta _{{\rm{dop}}}} = \left\lfloor {{{{N_{\rm{p}}}} \mathord{\left/
			{\vphantom {{{N_{\rm{p}}}} {{N_{\rm{g}}}}}} \right.
			\kern-\nulldelimiterspace} {{N_{\rm{g}}}}}} \right\rfloor $.
The difference between these beams is that ${\bf{p}}_0^{{M_{\rm{p}}}}$ should be replaced with ${\bf{p}}_{i{M_{\rm{g}}}}^{{M_{\rm{p}}}}$, or ${\bf{p}}_0^{{N_{\rm{p}}}}$ with ${\bf{p}}_{j{N_{\rm{g}}}}^{{N_{\rm{p}}}}$, where $1 \le i \le {\eta _{{\rm{del}}}} - 1$ and $1 \le j \le {\eta _{{\rm{dop}}}} - 1$. Note that at least one of the two pilot sequences needs to be replaced to ensure the orthogonality between columns. However, for the beam other than these ${\eta _{{\rm{del}}}}{\eta _{{\rm{dop}}}}$ beams, a new pair of pilot sequences should be used. And the coherence of the columns constructed by this new pair of pilot sequences and the columns constructed by other pairs of pilot sequences should be as low as possible to ensure the sensing performance of the entire pilot matrix, which is defined as the \emph{low coherence condition}.

To sum up, the designed deterministic pilots should satisfy both the \emph{orthogonality condition} and the \emph{low coherence condition}. ZC sequence is quite suitable for the proposed deterministic pilot design since it is a cyclic orthogonal sequence with good autocorrelation and low cross-correlation. We define the ZC sequence cyclically shifted by $c$ symbols with root $r$ and length $L$ as ${\bf{z}}_c^{L{\rm{,}}r}$, and its  $(k+1)$-th element can be expressed as
\begin{equation}
z_c^{L,r}\left[ k \right] = \frac{1}{{\sqrt L }}{e^{\frac{{{\bar \jmath}\pi r{{\left( {k - c} \right)}_L}\left( {{{\left( {k - c} \right)}_L} + {{\left( L \right)}_2}} \right)}}{L}}} ,
\end{equation}
where $k = 0,1, \cdots ,L - 1$. Note that $L$ in the current system should be ${{M_{\rm{p}}}}$ or ${{N_{\rm{p}}}}$. The cyclic orthogonality of the ZC sequence makes it satisfy the \emph{orthogonality condition}. Meanwhile, when the length of ZC sequence $L$ is prime, the magnitude of cross-correlation between two normalized ZC sequences with different roots is $\frac{1}{{\sqrt L }}$ \cite{1054840}. It means that when we need a new pair of pilot sequences, a pair of ZC sequences with a new pair of roots can be chosen, and they satisfy the \emph{low coherence condition}. Therefore, based on the ZC sequence, the proposed deterministic pilot design is summarized in {\bf Algorithm 1}, where  the pilots at all beams are rearranged as a matrix ${{\bf{\tilde X}}^{\rm{P}}}$ of size ${M_{\rm{p}}}{N_{\rm{p}}} \times {N_{\rm{t}}}$.

\IncMargin{0.2em}
\begin{algorithm}
	\SetAlgoNoLine 
	\SetKwInOut{Input}{\textbf{Input}}\SetKwInOut{Output}{\textbf{Output}} 
    \caption{Deterministic pilot design}
	\Input{
		${M_{\rm{p}}}$, ${N_{\rm{p}}}$, ${M_{\rm{g}}}$, ${N_{\rm{g}}}$\\
	}
	\Output{
		${{\bf{\tilde X}}^{\rm{P}}}$\\
	}
	\BlankLine
	
	Initialize: ${\eta _{{\rm{del}}}} = \left\lfloor {{{{M_{\rm{p}}}} \mathord{\left/
				{\vphantom {{{M_{\rm{p}}}} {{M_{\rm{g}}}}}} \right.
				\kern-\nulldelimiterspace} {{M_{\rm{g}}}}}} \right\rfloor$, 
${\eta _{{\rm{dop}}}} = \left\lfloor {{{{N_{\rm{p}}}} \mathord{\left/
			{\vphantom {{{N_{\rm{p}}}} {{N_{\rm{g}}}}}} \right.
			\kern-\nulldelimiterspace} {{N_{\rm{g}}}}}} \right\rfloor $, $g=0$, \\
	\qquad\qquad\ 	$b=0$, $\gamma  = {M_{\rm{p}}} - 1$, $\mu  = {N_{\rm{p}}} - 1$

	\While{$g{\eta _{{\rm{del}}}}{\eta _{{\rm{dop}}}} < {N_{\rm{t}}}$}{
		\For{$i=0$ to ${\eta _{{\rm{del}}}}-1$}{
			\For{$j=0$ to ${\eta _{{\rm{dop}}}}-1$}{
				$b = g{\eta _{{\rm{del}}}}{\eta _{{\rm{dop}}}} +  i {\eta _{{\rm{dop}}}} + j+1$ \\
				\If{$b > N_{\rm{t}}$}{
					break
				}
				${\left[ {{{{\bf{\tilde X}}}^{\rm{P}}}} \right]_{:,b}} = {\bf{z}}_{i{M_{\rm{g}}}}^{{M_{\rm{p}}}{\rm{,}}\gamma } \otimes {\bf{z}}_{j{N_{\rm{g}}}}^{{N_{\rm{p}}}{\rm{,}}\mu }$\\
			}
		}
		$g=g+1$, $\gamma = \gamma-1$, $\mu = \mu-1$
		
	}
	
	\Return ${{\bf{\tilde X}}^{\rm{P}}}$

\end{algorithm}

To analyze the correlation between the columns of the designed deterministic pilot matrix, we divide them into $g$ sets with  ${M_{\rm{g}}}{N_{\rm{g}}}{\eta _{{\rm{del}}}}{\eta _{{\rm{dop}}}}$ columns in each set.
The last set contains ${M_{\rm{g}}}{N_{\rm{g}}}{{\tilde \eta }_{{\rm{del}}}}{{\tilde \eta }_{{\rm{dop}}}}$ columns where ${{\tilde \eta }_{{\rm{del}}}} \le {\eta _{{\rm{del}}}}$ and ${{\tilde \eta }_{{\rm{dop}}}} \le {\eta _{{\rm{dop}}}}$. Hence the deterministic pilot matrix ${{\bf{X}}^{{\rm{DDB}}}}$ is given by
\begin{equation}\label{equ3-6}
\left[ {\overbrace {{{{\bf{\tilde c}}}_{1,1}}, \cdots {{{\bf{\tilde c}}}_{1,{M_{\rm{g}}}{N_{\rm{g}}}{\eta _{{\rm{del}}}}{\eta _{{\rm{dop}}}}}}}^{{\rm{group 1}}}, \cdots ,\overbrace {{{{\bf{\tilde c}}}_{g,1}}, \cdots {{{\bf{\tilde c}}}_{g,{M_{\rm{g}}}{N_{\rm{g}}}{{\tilde \eta }_{{\rm{del}}}}{{\tilde \eta }_{{\rm{dop}}}}}}}^{{\rm{group }}g}} \right].
\end{equation}
Thus the coherence between the columns of the pilot matrix ${{\bf{X}}^{{\rm{DDB}}}}$ can be given by
\begin{equation}\label{equ3-7}
\left| {{\bf{\tilde c}}_{{i_1},{j_1}}^{\rm{H}}{{{\bf{\tilde c}}}_{{i_2},{j_2}}}} \right| = \left\{ {\begin{array}{*{20}{l}}
	{1,}&{{i_1} = {i_2},{j_1} = {j_2}}\\
	{0,}&{{i_1} = {i_2},{j_1} \ne {j_2}}\\
	{\frac{1}{{\sqrt {{M_{\rm{p}}}{N_{\rm{p}}}} }},}&{{i_1} \ne {i_2}}
	\end{array}} \right..
\end{equation}

Note that the maximal coherence between the columns of the pilot matrix has a lower bound $\sqrt {\frac{{{M_{\rm{g}}}{N_{\rm{g}}}{N_{\rm{t}}} - {M_{\rm{p}}}{N_{\rm{p}}}}}{{\left( {{M_{\rm{g}}}{N_{\rm{g}}}{N_{\rm{t}}} - 1} \right){M_{\rm{p}}}{N_{\rm{p}}}}}} $, which is called Welch bound \cite{1055219}. In massive MIMO-OTFS systems, ${{M_{\rm{g}}}{N_{\rm{g}}}{N_{\rm{t}}}}$ is usually much larger than ${{M_{\rm{p}}}{N_{\rm{p}}}}$, which means that the maximal coherence  between the columns of the designed deterministic pilot matrix shown in (\ref{equ3-7}) is very close to Welch bound. Moreover, the designed deterministic pilot can be quickly generated according to different system configurations instead of spending a lot of memory for storage in advance, while the random pilot is hard to be implemented and consumes much memory due to its randomness.

Since the sensing matrix is the Hadamard product of pilot matrix ${{\bf{X}}^{{\rm{DDB}}}}$ and phase compensation matrix ${\bf{\Phi }}$ or ${\bf{\tilde \Phi }}$, the correlation between the columns of the sensing matrix is affected by the phase compensation matrix. Nevertheless, the proposed deterministic pilot design still has the advantage over the random pilot, as shown in simulation results.

\subsection{Pilot Overhead}
Comparing Fig. \ref{fig-new} with Fig. \ref{fig-ini}, we can find that
the delay-Doppler domain grids occupied by pilots and guard intervals  of the proposed pilot design 
is ${N_{\rm{g}}}\left( {{M_{\rm{p}}} + 2{M_{\rm{g}}}} \right)$ more than that of conventional random pilots used in \cite{shen2019channel} when the same number of pilots (i.e., ${M_{\rm{p}}}{N_{\rm{p}}}$) is considered. However, the proposed pilot design can save about 5-10 percent of the number of pilots compared with random pilots due to the better sensing performance, as shown in simulation results, which means that the overall pilot overhead of the proposed pilot design is less. Moreover, when $N$ is not large in practical systems, the fractional Doppler will cause  severe inter-Doppler interference, which means that pilots usually need to occupy the entire Doppler domain (i.e., ${N_{\rm{p}}} = N$) to prevent interference between the data and pilots. In this case, guard intervals along the Doppler-dimension are no longer required, and only the last ${{M_{\rm{g}}}}$ pilots are added to the beginning along the delay-dimension to construct the structure of Fig. \ref{fig-new}.
Therefore, the delay-Doppler domain grids occupied by pilots and guard intervals shown in Fig. \ref{fig-new} are the same as that of Fig. \ref{fig-ini}, which means that the overall saved pilot overhead can  further increase.

\subsection{MSMCE Algorithm}
In this subsection, we will give  a detailed discussion of the proposed MSMCE algorithm, which utilizes both two sensing matrices ${\bf{ \Theta }}$ and ${\bf{\tilde \Theta }}$ to solve the sparse signal recovery problem.

The delay-Doppler-beam domain channel ${{\bf{h}}^{{\rm{DDB}}}}$ can be recovered by the CS algorithm using the sensing matrix and received symbols. 
Usually, $N$ can not be too large in the practical system, which implies that the performance of the downlink CSI acquisition based on sensing matrix ${\bf{\tilde \Theta }}$ is unsatisfactory due to the inaccurate sensing matrix when the signal-to-noise ratio (SNR) is high.
Therefore, we prefer to use the sensing matrix ${\bf{ \Theta }}$, which is more accurate for channel estimation. However, according to (\ref{equ3-3}), the phase compensation matrix ${\bf{\Phi }}$ in the sensing matrix ${\bf{ \Theta }}$ depends on the information of each path. Therefore, when there is no knowledge of path delays and Doppler frequencies of the channel at first, we can choose ${\bf{\tilde \Theta }}$ as the sensing matrix to estimate the channel with CS algorithms since ${\bf{\tilde \Theta }}$ is irrelevant to the channel.  
Then we extract the path delay and Doppler frequency of each dominant path from the channel estimation results ${{\bf{\hat h}}^{{\rm{DDB}}}}$ to build a more accurate sensing matrix ${\bf{ \Theta }}$. Hence, before introducing the MSMCE algorithm, we propose a method to extract the path delays and Doppler frequencies from the channel estimation results, which is crucial in the MSMCE algorithm.

\begin{algorithm}
	\SetAlgoNoLine 
	\SetKwInOut{Input}{\textbf{Input}}\SetKwInOut{Output}{\textbf{Output}} 
	\caption{The path delay and Doppler frequency extraction}
	\Input{
		${{\bf{\hat h}}^{{\rm{DDB}}}}$, $\varepsilon$\\
	}
	\Output{
		$\mbox{Del}$, $\mbox{Dop}$\\
	}
	\BlankLine
	
	Initialize: $\mbox{Del} = \emptyset $, $\mbox{Dop} = \emptyset $
	
	Rearrange the magnitude of the channel estimation output ${{\bf{\hat h}}^{{\rm{DDB}}}}$ as ${\bf{H}}_{{\rm{mag}}}^{{\rm{DDB}}} \in {\mathbb{C}^{{N_{\rm{g}}} \times {M_{\rm{g}}} \times {N_{\rm{t}}}}}$.
	
	Sum the beam-dimension of the matrix ${\bf{H}}_{{\rm{mag}}}^{{\rm{DDB}}}$ to get  ${\bf{H}}_{{\rm{mag}}}^{{\rm{DD}}} \in {\mathbb{C}^{{N_{\rm{g}}} \times {M_{\rm{g}}}}}$.
	
	Sum the Doppler-dimension of the matrix ${\bf{H}}_{{\rm{mag}}}^{{\rm{DD}}}$ to get ${\bf{h}}_{{\rm{mag}}}^{\rm{D}} \in {\mathbb{C}^{{M_{\rm{g}}} \times 1 }}$.
	
	$\Sigma  = {\left\| {{\bf{h}}_{{\rm{mag}}}^{{\rm{D}}}} \right\|_1}$
	
	\While{${{{{\left\| {{\bf{h}}_{{\rm{mag}}}^{{\rm{D}}}} \right\|}_1}} \mathord{\left/
				{\vphantom {{{{\left\| {{\bf{H}}_{{\rm{mag}}}^{{\rm{DD}}}} \right\|}_1}} \Sigma }} \right.
				\kern-\nulldelimiterspace} \Sigma } > \varepsilon $ }{

		${l_\tau } = \mathop {\arg \max }\limits_{l \in \left\{ {1,2, \cdots ,{M_{\rm{g}}}} \right\}} {\left[ {{\bf{h}}_{{\rm{mag}}}^{{\rm{D}}}} \right]_l}$
		
		$\mbox{Del} = \left\{ {\mbox{Del},{l_\tau }} \right\}$

		${k_{{\nu _1}}} = \mathop {{\mathop{\rm argmax}\nolimits} }\limits_{k \in \left\{ {1,2, \cdots ,{N_{\rm{g}}}} \right\}} {\left[ {{\bf{H}}_{{\rm{mag}}}^{{\rm{DD}}}} \right]_{k,{l_\tau }}}$

		${k_{{\nu _2}}} = \mathop {{\mathop{\rm argmax}\nolimits} }\limits_{k \in \left\{ {1,2, \cdots ,{N_{\rm{g}}}} \right\}\backslash \left\{ {{k_{{\nu _1}}}} \right\}} {\left[ {{\bf{H}}_{{\rm{mag}}}^{{\rm{DD}}}} \right]_{k,{l_\tau }}}$
		
		${k_\nu } \!=\! {k_{{\nu _1}}} - \left\lfloor {\frac{{{N_{\rm{g}}}}}{2}} \right\rfloor  - 1,\  {\tilde k_\nu } \!=\! \frac{{\left( {{k_{{\nu _2}}} - {k_{{\nu _1}}}} \right){{\left[ {{\bf{H}}_{{\rm{mag}}}^{{\rm{DD}}}} \right]}_{{k_{{\nu _2}}},{l_\tau }}}}}{{{{\left[ {{\bf{H}}_{{\rm{mag}}}^{{\rm{DD}}}} \right]}_{{k_{{\nu _1}}},{l_\tau }}} + {{\left[ {{\bf{H}}_{{\rm{mag}}}^{{\rm{DD}}}} \right]}_{{k_{{\nu _2}}},{l_\tau }}}}}$
		
		$\mbox{Dop} = \left\{ {\mbox{Dop},{k_\nu } + {{\tilde k}_\nu }} \right\}$
		
		Set the $l_\tau$-th element of ${\bf{h}}_{{\rm{mag}}}^{\rm{D}}$ to 0.

	}

	\Return $\mbox{Del}$, $\mbox{Dop}$
	
\end{algorithm}

We first define the magnitude of the delay-Doppler-beam domain channel as $H_{{\rm{mag}}}^{{\rm{DDB}}}\left[ {k,{l},b} \right] \buildrel \Delta \over = \left| {{H^{{\rm{DDB}}}}\left[ {k,l,b} \right]} \right|$. The delay of each path can be easily obtained by summing the Doppler-dimension and the beam-dimension of $H_{{\rm{mag}}}^{{\rm{DDB}}}\left[ {k,{l},b} \right]$ and searching positions with dominant values along the delay-dimension. Hence, we focus on the extraction of the Doppler frequency ${\nu _i} = \frac{{{k_i} + {{\tilde k}_i}}}{{N{T_{{\rm{sym}}}}}}$ of each dominant path. Note that we approximate the Doppler frequency of all subpaths of each dominant path to the same. Therefore, for the $(i+1)$-th dominant path, $H_{{\rm{mag}}}^{{\rm{DDB}}}\left[ {k,{l_i},b} \right]$ can be expressed as
\begin{align}
&H_{{\rm{mag}}}^{{\rm{DDB}}}\left[ {k,{l_i},b} \right] \nonumber \\
&\quad\approxeq \!\sqrt {\frac{N}{{{N_{\rm{t}}}}}}\left| {\sum\limits_{{s_i} = 0}^{S - 1} {{h_{{s_i}}}{e^{{\bar\jmath}2\pi \left( {{M_{{\rm{CP}}}} - {l_i}} \right){T_{\rm{s}}}{\nu _{{s_i}}}}}} {\Xi _{{N_{\rm{t}}}}}\!\left( {b - {{{N_{\rm{t}}}\sin {\varphi _{{s_i}}}} \mathord{\left/
				{\vphantom {{{N_{\rm{t}}}\sin {\varphi _{{s_i}}}} 2}} \right.
				\kern-\nulldelimiterspace} 2}} \right)} \right|\nonumber \\
&\qquad\qquad\qquad \times \left| {{\Xi _N}\left( {k - N{T_{{\rm{sym}}}}{\nu _i}} \right)} \right| .
\end{align}
Then we sum the beam-dimension of $H_{{\rm{mag}}}^{{\rm{DDB}}}\left[ {k,{l_i},b} \right]$ as
\begin{align}\label{equ-HDD}
H_{{\rm{mag}}}^{{\rm{DD}}}\left[ {k,{l_i}} \right] &= \sum\limits_{b =  - {{{N_{\rm{t}}}} \mathord{\left/
			{\vphantom {{{N_{\rm{t}}}} 2}} \right.
			\kern-\nulldelimiterspace} 2}}^{{{{N_{\rm{t}}}} \mathord{\left/
			{\vphantom {{{N_{\rm{t}}}} 2}} \right.
			\kern-\nulldelimiterspace} 2} - 1} {H_{{\rm{mag}}}^{{\rm{DDB}}}\left[ {k,{l_i},b} \right]} \nonumber \\
		& \approxeq \beta_i \left| {{\Xi _N}\left( {k - N{T_{{\rm{sym}}}}{\nu _i}} \right)} \right|,
\end{align}
where
\begin{align}
&{\beta _i} = \sqrt {\frac{N}{{{N_{\rm{t}}}}}} \nonumber \\
&\quad\times \!\!\!\!\sum\limits_{b =  - {{{N_{\rm{t}}}} \mathord{\left/
			{\vphantom {{{N_{\rm{t}}}} 2}} \right.
			\kern-\nulldelimiterspace} 2}}^{{{{N_{\rm{t}}}} \mathord{\left/
			{\vphantom {{{N_{\rm{t}}}} 2}} \right.
			\kern-\nulldelimiterspace} 2} - 1} \!{\left| {\sum\limits_{{s_i} = 0}^{S - 1} \!{{h_{{s_i}}}{e^{{\bar\jmath}2\pi \left( {{M_{{\rm{CP}}}} - {l_i}} \right){T_{\rm{s}}}{\nu _{{s_i}}}}}} {\Xi _{{N_{\rm{t}}}}}\!\left( {b\! - \!{{{N_{\rm{t}}}\sin {\varphi _{{s_i}}}} \mathord{\left/
					{\vphantom {{{N_{\rm{t}}}\sin {\varphi _{{s_i}}}} 2}} \right.
					\kern-\nulldelimiterspace} 2}} \right)} \right|}  .
\end{align}
Note that $\left| {{\Xi _N}\left( {k - N{T_{{\rm{sym}}}}{\nu _i}} \right)} \right|$ in (\ref{equ-HDD}) reaches the maximum at $k = {k_i} + {\tilde k_i}$ and decreases as $k$ moves away from ${k_i} + {\tilde k_i}$. Hence, we define that
\begin{align}
&{k_{{\nu _1}}} = \mathop {{\mathop{\rm argmax}\nolimits} }\limits_{k \in \left\{ {\left\lceil {{{ - {N_{\rm{g}}}} \mathord{\left/
					{\vphantom {{ - {N_{\rm{g}}}} 2}} \right.
					\kern-\nulldelimiterspace} 2}} \right\rceil , \cdots ,\left\lceil {{{{N_{\rm{g}}}} \mathord{\left/
					{\vphantom {{{N_{\rm{g}}}} 2}} \right.
					\kern-\nulldelimiterspace} 2}} \right\rceil  - 1} \right\}} H_{{\rm{mag}}}^{{\rm{DD}}}\left[ {k,{l_i}} \right], \nonumber \\
& {k_{{\nu _2}}} = \mathop {{\mathop{\rm argmax}\nolimits} }\limits_{k \in \left\{ {\left\lceil {{{ - {N_{\rm{g}}}} \mathord{\left/
									{\vphantom {{ - {N_{\rm{g}}}} 2}} \right.
									\kern-\nulldelimiterspace} 2}} \right\rceil , \cdots ,\left\lceil {{{{N_{\rm{g}}}} \mathord{\left/
									{\vphantom {{{N_{\rm{g}}}} 2}} \right.
									\kern-\nulldelimiterspace} 2}} \right\rceil  - 1} \right\}\backslash \left\{ {{k_{{\nu _1}}}} \right\}} H_{{\rm{mag}}}^{{\rm{DD}}}\left[ {k,{l_i}} \right].
\end{align}
It can be checked that the integer part of the Doppler frequency is ${k_i} = {k_{{\nu _1}}}$, $\left| {{k_{{\nu _2}}} - {k_{{\nu _1}}}} \right| = 1$, and ${k_i} + {\tilde k_i}$ must be between ${k_{{\nu _1}}}$ and ${k_{{\nu _2}}}$. Then we can get that
\begin{align}\label{equ-fractional}
\frac{{H_{{\rm{mag}}}^{{\rm{DD}}}\left[ {{k_{{\nu _1}}},{l_i}} \right]}}{{H_{{\rm{mag}}}^{{\rm{DD}}}\left[ {{k_{{\nu _2}}},{l_i}} \right]}}&\!=\! \frac{{\left| {{\Xi _N}( - {{\tilde k}_i})} \right|}}{{\left| {{\Xi _N}({k_{{\nu _2}}} - {k_{{\nu _1}}} - {{\tilde k}_i})} \right|}}\nonumber \\
&\!\mathop  = \limits^{(\rm{a})}\! \left| {\frac{{\sin \left( { - {{\tilde k}_i}\pi } \right)}}{{\sin \left( {{\textstyle{{ - {{\tilde k}_i}} \over N}}\pi } \right)}}} \right|\!\! \cdot\!\! {\left| {\frac{{\sin\! \left(\! {\left( {{k_{{\nu _2}}} - {k_{{\nu _1}}} - {{\tilde k}_i}} \right)\!\pi } \!\right)}}{{\sin \left( {{\textstyle{{\left( {{k_{{\nu _2}}} - {k_{{\nu _1}}} - {{\tilde k}_i}} \right)} \over N}}\pi } \right)}}} \right|^{ - 1}},
\end{align}
where (a) follows from the fact that $\left| {{\Xi _N}\left( x \right)} \right| = \frac{1}{N}\left| {\frac{{\sin \left( {\pi x} \right)}}{{\sin \left( {\frac{{\pi x}}{N}} \right)}}} \right|$. Since $\left| {{k_{{\nu _2}}} - {k_{{\nu _1}}}} \right| = 1$, which means that $\left| {\sin \left( { - {{\tilde k}_i}\pi } \right)} \right| = \left| {\sin \left( {\left( {{k_{{\nu _2}}} - {k_{{\nu _1}}} - {{\tilde k}_i}} \right)\pi } \right)} \right|$, (\ref{equ-fractional}) can be rewritten as
\begin{align}\label{equ-fra2}
\frac{{H_{{\rm{mag}}}^{{\rm{DD}}}\left[ {{k_{{\nu _1}}},{l_i}} \right]}}{{H_{{\rm{mag}}}^{{\rm{DD}}}\left[ {{k_{{\nu _2}}},{l_i}} \right]}} &= \frac{{\left| {\sin \left( {{\textstyle{{\left( {{k_{{\nu _2}}} - {k_{{\nu _1}}} - {{\tilde k}_i}} \right)} \over N}}\pi } \right)} \right|}}{{\left| {\sin \left( {{\textstyle{{ - {{\tilde k}_i}} \over N}}\pi } \right)} \right|}} \nonumber \\
&\approx \frac{{\left| {\left( {{k_{{\nu _2}}} - {k_{{\nu _1}}} - {{\tilde k}_i}} \right)} \right|}}{{\left| { - {{\tilde k}_i}} \right|}}.
\end{align}
Therefore, the fractional Doppler can be derived from (\ref{equ-fra2}) as
\begin{equation}
{\tilde k_i} = \frac{{\left( {{k_{{\nu _2}}} - {k_{{\nu _1}}}} \right)H_{{\rm{mag}}}^{{\rm{DD}}}\left[ {{k_{{\nu _2}}},{l_i}} \right]}}{{H_{{\rm{mag}}}^{{\rm{DD}}}\left[ {{k_{{\nu _1}}},{l_i}} \right] + H_{{\rm{mag}}}^{{\rm{DD}}}\left[ {{k_{{\nu _2}}},{l_i}} \right]}}.
\end{equation}
Similarly, the tap index set of the path delay $\mbox{Del}$ and Doppler frequency $\mbox{Dop}$ of all dominant paths are obtained, and the process is summarized in {\bf Algorithm 2}, where $\varepsilon$ is used to terminate the iteration and usually a smaller value. Note that there is no interference between the dominant paths in Algorithm 2, which is because that the fractional delay is not considered in this paper.

Next, the proposed MSMCE algorithm is given in {\bf Algorithm 3}. If we have no information about the channel to be estimated at the beginning,  ${\bf{\tilde \Theta }}$ should be used as the sensing matrix to obtain a temporary channel estimation result by the CS algorithm. Then we use the initial estimation result to extract the path delays and Doppler frequencies of all dominant paths, which are used to modify the inaccurate sensing matrix to obtain a more accurate sensing matrix ${\bf{\Theta }}$. Finally, the channel estimation result ${{\bf{\hat h}}^{{\rm{DDB}}}}$ is obtained by the CS algorithm based on ${\bf{\Theta }}$, and $\mbox{Del}$ and $\mbox{Dop}$ are also updated for the next channel estimation. Since the channel geometry changes very slowly relative to the communication timescale \cite{monk2016otfs}, in the next channel estimation, $\mbox{Del}$ and $\mbox{Dop}$ can be directly used to construct the modified sensing matrix to estimate the channel.

\begin{algorithm}
	\SetAlgoNoLine 
	\SetKwInOut{Input}{\textbf{Input}}\SetKwInOut{Output}{\textbf{Output}} 
	\caption{MSMCE Algorithm}
	\Input{
		${{\bf{X}}^{{\rm{DDB}}}}$, ${{\bf{y}}^{{\rm{DD}}}}$, $\mbox{Del}$, $\mbox{Dop}$\\
	}
	\Output{
		${{\bf{\hat h}}^{{\rm{DDB}}}}$, $\mbox{Del}$, $\mbox{Dop}$\\
	}
	\BlankLine

	\If{ $\rm{Del} = \emptyset $ or  $\rm{Dop} = \emptyset $}{
		Sensing martix ${\bf{\tilde \Theta }} = {\bf{\tilde \Phi }} \odot {{\bf{X}}^{{\rm{DDB}}}}$
		
		Calculate the channel estimation result ${{\bf{\hat h}}^{{\rm{DDB}}}}$ by the CS algorithm based on ${\bf{\tilde \Theta }}$ and ${{\bf{y}}^{{\rm{DD}}}}$.
		
		Calculate the $\mbox{Del}$ and $\mbox{Dop}$ by the Algorithm 2 based on ${{\bf{\hat h}}^{{\rm{DDB}}}}$.	
	}
	
	Use $\mbox{Del}$ and $\mbox{Dop}$ to construct the phase compensation matrix ${\bf{\Phi }}$.
	
	Sensing martix ${\bf{ \Theta }} = {\bf{\Phi }} \odot {{\bf{X}}^{{\rm{DDB}}}}$
	
	Calculate the channel estimation result ${{\bf{\hat h}}^{{\rm{DDB}}}}$ by the CS algorithm based on ${\bf{\Theta }}$ and ${{\bf{y}}^{{\rm{DD}}}}$.
	
	Update the $\mbox{Del}$ and $\mbox{Dop}$ by the Algorithm 2 based on ${{\bf{\hat h}}^{{\rm{DDB}}}}$.	
	
	\Return ${{\bf{\hat h}}^{{\rm{DDB}}}}$, $\mbox{Del}$, $\mbox{Dop}$
	
\end{algorithm}

\subsection{The Performance Analysis of the MSMCE Algorithm}
The complexity of the proposed MSMCE algorithm mainly comes from two aspects: the Algorithm 2 and the selected CS algorithm. 
We can find that most of the operations in the iteration process of Algorithm 2 are comparisons, the number of which is about $P\left( {{M_{\rm{g}}} + 2{N_{\rm{g}}}}-1 \right)$, where $P$ is the number of the dominant paths.
Therefore, the complexity of Algorithm 2 is very low compared with the  CS algorithm, which means that the MSMCE algorithm can obtain a significant performance gain at the cost of a negligible increase in the complexity.

However, note that the MSMCE algorithm is based on the assumption that we approximate the Doppler frequency of all subpaths of each dominant path to the same value ${\nu _i} = \frac{{{k_i} + {{\tilde k}_i}}}{{N{T_{{\rm{sym}}}}}}$. We now discuss the influence of the approximation error caused by this assumption on the proposed MSMCE algorithm.

Recalling (\ref{equ2-16}), the approximation error can be defined as follows
\begin{equation} \label{equ-error}
\zeta  = \sum\limits_{i = 0}^{P - 1} {\sum\limits_{{s_i} = 0}^{S - 1} {\left| {{e^{{\bar\jmath}2\pi \frac{{l\left( {{k_{{s_i}}} + {{\tilde k}_{{s_i}}} - {k_i} - {{\tilde k}_i}} \right)}}{{\left( {M + {M_{{\rm{CP}}}}} \right)N}}}} - 1} \right|} } .
\end{equation}
We can find that if each dominant path contains only one subpath, there is no approximation error, i.e., $\zeta  = 0$, and the MSMCE algorithm achieves the optimal performance. However, if more subpaths are considered, $\zeta$ will increase and lead to a decrease in the accuracy of the derived downlink CSI acquisition model and the extracted path information, and eventually make the performance of the MSMCE algorithm degrades.

Such an approximation error is related to several factors. First, the user velocity will affect $\zeta$. This is because that although the directions of arrival deviation of subpaths are very slight, too high the velocity will still make the Doppler frequency difference between subpaths become larger and increase the approximation error. Second, it is noticed that $l$ in (\ref{equ-error}) is the position of pilots along the delay-dimension, which implies that the pilot position will have an influence on the channel estimation performance. Specifically, the larger the $l$, the more severe the approximation error. Therefore, the initial pilot position along the delay-dimension $l_{\rm{p}}$ should be as close to zero as possible to obtain an accurate CSI. Note that the pilot position along the Doppler-dimension will not affect the performance since it is irrelevant to the approximation error shown in (\ref{equ-error}).

Moreover, for the proposed MSMCE algorithm, there are two special cases that we need to discuss. First, when  $N$ is large enough that the resolution of the Doppler domain is sufficient to map the Doppler frequency of each path to an integer tap, the model described in (\ref{equ2-25}) is accurate in this case, and the phase compensation matrix ${\bf{\tilde \Phi }}$ is exact. Therefore, there is no need to extract the path delays and Doppler frequencies from the previous channel estimation results to modify the sensing matrix, and only steps 2-3 in the MSMCE algorithm are needed to estimate the channel. 
Second, when the channel geometry changes very rapidly or the time intervals between the channel estimations are too long, the path information $\mbox{Del}$ and $\mbox{Dop}$ extracted from the previous channel estimation results is not accurate enough, which leads to the degradation of the MSMCE algorithm performance. In this case, we need to re-extract the path information with the current channel estimation result and use the extracted information to construct the modified sensing matrix to obtain a more accurate CSI.

\section{Simulation Results}
In this section, we utilize the numerical simulation to illustrate the performance of the proposed downlink CSI acquisition scheme. We choose the 3D-SOMP algorithm proposed in \cite{shen2019channel} as the CS algorithm used in MSMCE. The performance of the initial sensing matrix based channel estimation (ISMCE) is also presented for comparison, where only ${\bf{\tilde \Theta }}$ is used as the sensing matrix (i.e., the steps 2-3 in the MSMCE algorithm).
The normalized mean square error (NMSE) is computed as
\begin{equation}\label{equ4-1}
{\rm{NMSE = }}\frac{{\left\| {{{{\bf{\hat h}}}^{{\rm{DDB}}}} - {{\bf{h}}^{{\rm{DDB}}}}} \right\|_2^2}}{{\left\| {{{\bf{h}}^{{\rm{DDB}}}}} \right\|_2^2}}.
\end{equation}
We use the quasi deterministic radio channel generator (QuaDRiGa) to generate over $10^5$ channel realizations for the Monte-Carlo simulations \cite{Quadriga}. The relevant simulation parameters are given in Table \ref{tab1}.
Note that since $N$ is not large, the sparsity of the delay-Doppler-beam domain channel along the Doppler-dimension is not obvious. Therefore, we set $N_{\rm{P}} = N$ to prevent the interference between the data and pilots. Unless explicitly mentioned otherwise, the pilots used are the proposed deterministic pilots. The pilot overhead ratio is defined as the ratio between the number of pilots and total delay-Doppler domain resource grids, i.e., $\frac{{{M_{\rm{p}}}{N_{\rm{p}}}}}{{MN}}$.

\begin{table}
	\caption{Simulation parameters}
	\label{table}
	\centering
	\setlength{\tabcolsep}{10pt}
	\begin{tabular}{|c|c|}
		\hline
		Parameter& 	Values \\
		\hline
		Carrier frequency (GHz) & 4 \\
		\hline
		Subcarrier spacing (kHz)& 15 \\
		\hline
		Size of OTFS symbol $\left( M, N\right)$ & (512, 19 / 11 / 7) \\
		\hline
		Length of CP $M_{\rm{CP}}$ & 128 \\
		\hline
		Number of BS antennas & 16\ /\ 32\ /\ 64 \\
		\hline
		Number of user terminal antennas & 1\\
		\hline
		Scenario & Urban macro cell \\
		\hline
		The number of dominant paths & 6\\
		\hline
		The number of subpaths per dominant path & 1\ /\ 5\ /\ 10\ /\ 20\\
		\hline
		User velocity (m/s) & 10 $\sim$ 130 \\
		\hline
		
	\end{tabular}
	\label{tab1}
\end{table}

\begin{figure}
	\centering
	\includegraphics[width=0.5\textwidth]{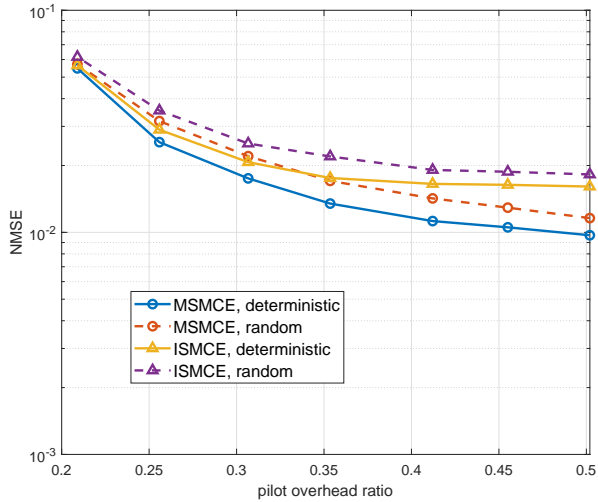}
	\caption{The NMSE performance versus pilot overhead ratio. The number of BS antennas is 16, the SNR is 5 dB, and the user velocity is 100 m/s.}
	\label{fig-pilot_overhead}
\end{figure}
In Fig. \ref{fig-pilot_overhead}, we compare the NMSE performance of different algorithms with different types of pilots against the pilot overhead ratio. The random pilots consist of complex Gaussian random sequences. The number of BS antennas is 16, the SNR is 5dB, $N=19$, the user velocity is 100 m/s, and the number of subpaths per dominant path is 20. We observe that the proposed MSMCE algorithm outperforms the ISMCE algorithm under different pilot overhead ratios when the same type of pilots is utilized. This is because that the MSMCE algorithm considers the fractional Doppler and uses the information extracted from the channel to reconstruct a more accurate sensing matrix. Moreover, the proposed deterministic pilot design has a better performance than the conventional random pilots when the same algorithm is used. For example, when the MSMCE algorithm is used, only 35\% pilot overhead ratio is required for the proposed deterministic pilot design to achieve the NMSE of 0.013, while random pilots need at least 45\% pilot overhead ratio.  This is due to the fact that the coherence between the columns of the designed pilot matrix is very low and has a better sensing performance compared with random pilots.

\begin{figure}
	\centering
	\includegraphics[width=0.5\textwidth]{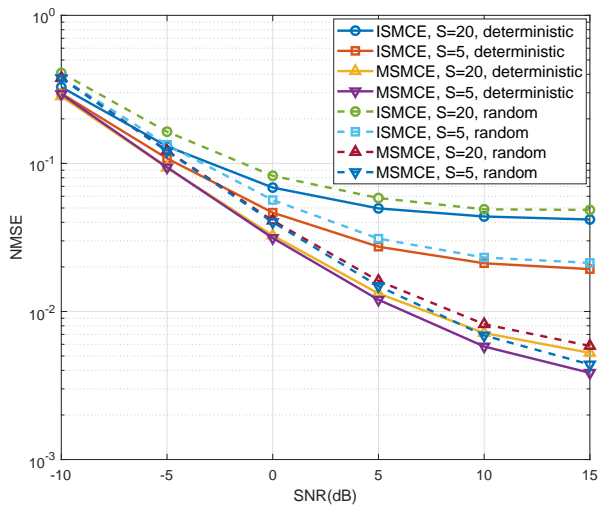}
	\caption{The NMSE performance versus SNR and comparison of the MSMCE algorithm and the ISMCE algorithm. The number of BS antennas is 64, and the user velocity is 100 m/s.}
	\label{fig2}
\end{figure}

Fig. \ref{fig2} shows the NMSE performance comparison of the MSMCE algorithm and the ISMCE algorithm against the SNR, where the number of subpaths per dominant path (i.e., $S$) is 20 or 5, and the pilots are random or deterministic. The number of BS antennas is 64, the user velocity is 100 m/s, $N=19$, and the pilot overhead ratio is 89\%. Such a high pilot overhead is because the beam domain of the generated channel is not pure sparse when the DFT matrix is used to convert the channel from the space domain to the beam domain\cite{7841676}, which means that more pilots are needed for both the ISMCE algorithm and the MSMCE algorithm.
We observe that our proposed MSMCE algorithm has a substantial performance gain over the ISMCE algorithm, regardless of the number of subpaths. This is due to the fact that the MSMCE algorithm uses the extracted path information to modify the sensing matrix to obtain more accurate CSI. Moreover, the channel estimation with the proposed deterministic pilots outperforms that with the random pilots due to the better sensing performance.

\begin{figure}
	\centering
	\includegraphics[width=0.5\textwidth]{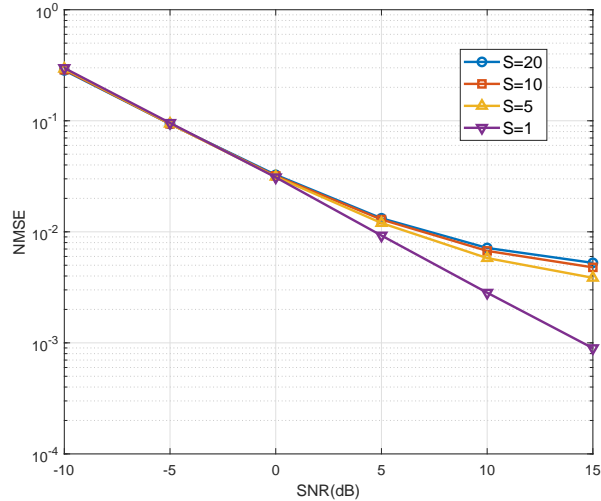}
	\caption{The NMSE performance versus SNR and comparison of the MSMCE algorithm with different numbers of subpaths per dominant path. The number of BS antennas is 64, the user velocity is 100 m/s.}
	\label{fig3}
\end{figure}

Fig. \ref{fig3} presents the NMSE performance comparison of the MSMCE algorithm with different numbers of subpaths per dominant path. The number of BS antennas is 64, the user velocity is 100 m/s, $N=19$, and the pilot overhead ratio is 89\%. It can be checked that with the decrease in the number of subpaths per dominant path, the performance of the MSMCE algorithm improves, and the optimal performance is obtained when the number of subpaths is 1. This is because that the less the number of subpaths, the smaller the approximation error shown in (\ref{equ-error}). Moreover, the sensing matrix ${\bf{\Theta }}$ is completely accurate when each dominant path consists of only one subpath since there is no approximation error.

\begin{figure}
	\centering
	\includegraphics[width=0.5\textwidth]{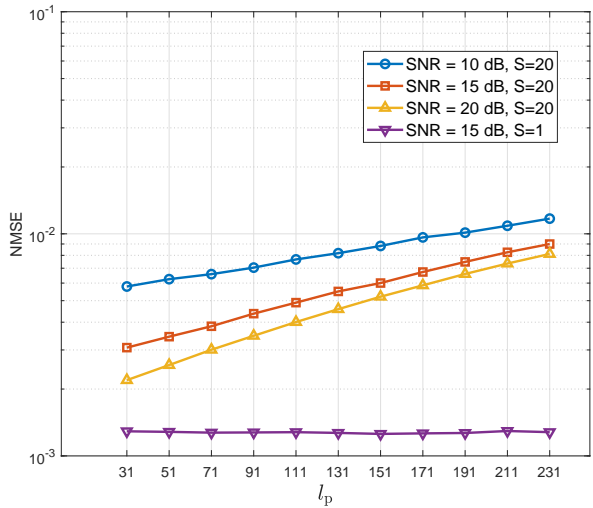}
	\caption{The NMSE performance versus $l_{\rm{p}}$ and comparison of the MSMCE algorithm under different SNRs. The number of BS antennas is 32, and the user velocity is 100 m/s.}
	\label{fig4}
\end{figure}

In Fig. \ref{fig4}, we show the NMSE performance of the MSMCE algorithm with different initial positions of pilots along the delay-dimension, i.e., $l_{\rm{p}}$. The number of BS antennas is 32, the user velocity is 100 m/s, $N=19$, and the pilot overhead ratio is 49\%. We observe that the performance of the MSMCE algorithm degrades with an increase of $l_{\rm{p}}$ when the number of subpaths per dominant path is 20. 
This is because that the approximation error increases with the $l_{\rm{p}}$ and makes the performance of the channel estimation degrades.
We can also observe that the slope of the curves increases with the increment of SNR, which means that the impact of the approximation error is more obvious in the high SNR regime. 
However, when each dominant path contains only one subpath, the performance is irrelevant to the position of pilots since there is no approximation error in this case.

\begin{figure}
	\centering
	\includegraphics[width=0.5\textwidth]{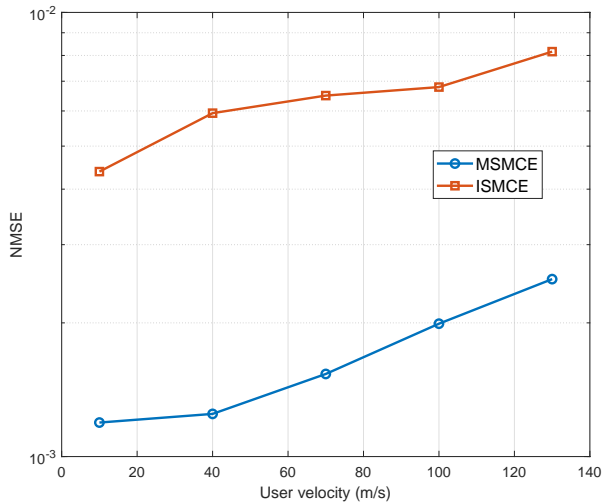}
	\caption{The NMSE performance versus user velocity. The number of BS antennas is 16, and the SNR is 15 dB.}
	\label{fig-velocity}
\end{figure}

Fig. \ref{fig-velocity} shows the NMSE performance comparison at different user velocities. The number of BS antennas is 16, the SNR is 15 dB, $N=19$, the pilot overhead ratio is 35\%, and the number of subpaths per dominant path is 20. 
We observe that the NMSE performance of the MSMCE algorithm degrades as the user velocity increases. This is because that the difference between the Doppler frequencies of subpaths increases with the user velocity, which leads to an increase in the approximation error. However, the MSMCE algorithm still has a significant performance gain over the ISMCE algorithm at different user velocities.

\begin{figure}
	\centering
	\includegraphics[width=0.5\textwidth]{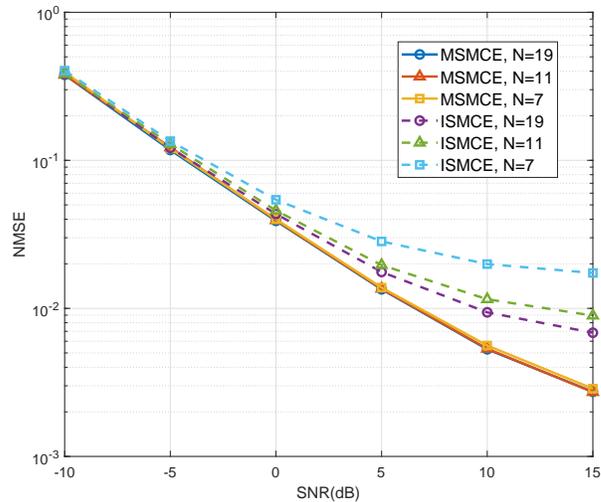}
	\caption{The NMSE performance versus SNR and comparison of the MSMCE algorithm and the ISMCE algorithm  under different OTFS symbol sizes. The number of BS antennas is 16, and the user velocity is 100 m/s.}
	\label{fig-N}
\end{figure}

In Fig. \ref{fig-N}, we show the NMSE performance comparison under different OTFS symbol sizes. The number of BS antennas is 16, the user velocity is 100 m/s, the pilot overhead ratio is 35\%, and the number of subpaths per dominant path is 20. We can observe that as $N$ decreases, the performance of the ISMCE algorithm degrades, while the performance of the MSMCE algorithm remains consistent. This is because that the decrease in $N$ will further reduce the resolution of the Doppler domain, which makes the influence of the fractional Doppler more serious. Therefore, the ISMCE algorithm performs worse under lower $N$ due to the omission of fractional Doppler.

\begin{figure}
	\centering
	\includegraphics[width=0.5\textwidth]{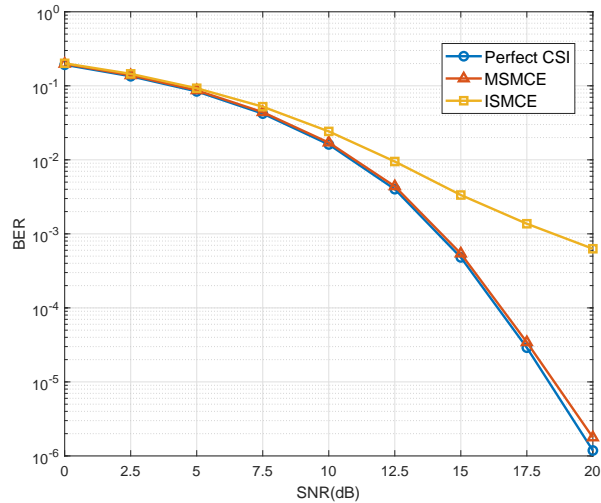}
	\caption{The BER performance versus SNR. The user velocity is 100 m/s, and the pilot overhead ratio is 35\%.}
	\label{fig-ber}
\end{figure}

Finally, Fig. \ref{fig-ber} shows the bit-error-rate (BER) performance comparison of different channel estimation algorithms. The transmitted symbols in the delay-Doppler domain are modulated by 4QAM. The user velocity is 100 m/s, $N=11$, the pilot overhead ratio is 35\%, and the number of subpaths per dominant path is 1. The minimum mean square error (MMSE) detector is used to recover the transmitted data with the CSI obtained by the MSMCE algorithm and the ISMCE algorithm. Moreover, the BER performance of the perfect CSI is provided for comparison. We can observe that the proposed MSMCE algorithm can achieve satisfying BER performance, which outperforms the ISMCE algorithm and is very close to the perfect CSI case due to the consideration of the fractional Doppler and the utilization of the extracted path information.

\section{Conclusion}
In this paper, we proposed a downlink CSI acquisition scheme for massive MIMO-OTFS systems in presence of the fractional Doppler, including deterministic pilot design and channel estimation algorithm. Considering the fractional Doppler in practical systems, we first analyzed the input-output relationship of SISO OTFS systems based on the OFDM modem and extended it to massive MIMO-OTFS systems. Next, based on the downlink CSI acquisition model, the deterministic pilot was designed, and the MSMCE algorithm was presented to reconstruct the delay-Doppler-beam domain channel. Simulation results demonstrated that the proposed scheme could acquire accurate downlink CSI.

\appendix[Proof of Proposition 1]

By substituting (\ref{equ2-9}) into (\ref{equ2-15}), ${{Y^{{\rm{DD}}}}\left[ {k,l} \right]}$ can be expressed as
\begin{align}\label{equA-1}
{Y^{{\rm{DD}}}}\left[ {k,l} \right] &= \frac{1}{{NM}}\sum\limits_{k' = \left\lceil { - {N \mathord{\left/
				{\vphantom {N 2}} \right.
				\kern-\nulldelimiterspace} 2}} \right\rceil }^{\left\lceil {{N \mathord{\left/
				{\vphantom {N 2}} \right.
				\kern-\nulldelimiterspace} 2}} \right\rceil  - 1} \sum\limits_{l' = 0}^{M - 1} {\alpha _{k,l}}\left[ {k',l'} \right] \nonumber \\
		&\qquad\qquad\qquad\times{{X^{{\rm{DD}}}}\left[ {k',l'} \right]}  + {Z^{{\rm{DD}}}}\left[ {k,l} \right] ,
\end{align}
where ${{\alpha _{k,l}}\left[ {k',l'} \right]}$ is given by 
\begin{align}\label{equA-2}
&{\alpha _{k,l}}\left[ {k',l'} \right]  \nonumber\\
&= \sum\limits_{n = 0}^{N - 1} \sum\limits_{m = 0}^{M - 1} \sum\limits_{m' = 0}^{M - 1} {\rm{ }}{e^{{\bar\jmath}2\pi \frac{{\left( {m - m'} \right){M_{{\rm{CP}}}}}}{M}}}{e^{ - {\bar\jmath}2\pi \frac{{n\left( {k - k'} \right)}}{N}}}{e^{{\bar\jmath}2\pi \frac{{ml - m'l'}}{M}}}   \nonumber\\
&\qquad\times   \iint {{h_{\rm{c}}}(\tau ,\nu ){A_{{g_{{\rm{rx}}}},{g_{{\rm{tx}}}}}}( - \tau ,(m - m')\Delta f - \nu )} {e^{ - {\bar\jmath}2\pi \nu \tau }}\nonumber \\
&\qquad \times{e^{ - {\bar\jmath}2\pi \left( {m'\Delta f\tau  - n{T_{{\rm{sym}}}}\nu } \right)}}d\tau d\nu   \nonumber \\
&= \sum\limits_{n = 0}^{N - 1} \sum\limits_{m = 0}^{M - 1} \sum\limits_{m' = 0}^{M - 1}{\rm{ }}{e^{{\bar\jmath}2\pi \frac{{\left( {m - m'} \right){M_{{\rm{CP}}}}}}{M}}}{e^{ - {\bar\jmath}2\pi \frac{{n\left( {k - k'} \right)}}{N}}}{e^{{\bar\jmath}2\pi \frac{{ml - m'l'}}{M}}}  \nonumber \\
& \qquad\times   \sum\limits_{i = 0}^{P - 1} \sum\limits_{{s_i} = 0}^{S - 1} {{h_{{s_i}}}} \frac{1}{M}\sum\limits_{p = {M_{{\rm{CP}}}} - {l_i}}^{M + {M_{{\rm{CP}}}} - 1 - {l_i}}\!\! {e^{ - {\bar\jmath}2\pi \left( {m'\Delta f{\tau _i} - n{T_{{\rm{sym}}}}{\nu _{{s_i}}}} \right)}} \nonumber \\
&\qquad \times{e^{ - {\bar\jmath}2\pi {\nu _{{s_i}}}{\tau _i}}}{{e^{ - {\bar\jmath}2\pi \left( {(m - m')\Delta f - {\nu _{{s_i}}}} \right)(\frac{p}{{M\Delta f}} + {\tau _i})}}}  \! \nonumber \\
&= N\sum\limits_{i = 0}^{P - 1} {\sum\limits_{{s_i} = 0}^{S - 1} {{h_{{s_i}}}} } {\Xi _N}\left( {k - k' - {k_{{s_i}}} - {{\tilde k}_{{s_i}}}} \right) \nonumber \\
& \qquad\times  M\sum\limits_{p = {M_{{\rm{CP}}}} - {l_i}}^{M + {M_{{\rm{CP}}}} - 1 - {l_i}} \!{{e^{{\bar\jmath}2\pi \frac{{p\left( {{k_{{s_i}}} + {{\tilde k}_{{s_i}}}} \right)}}{{\left( {M + {M_{\rm{CP}}}} \right)N}}}}} \nonumber \\
& \qquad\times\delta \left( {{{\left( {p - {M_{{\rm{CP}}}} + {l_i} - l} \right)}_M}} \right)\delta \left( {{{\left( {p - {M_{{\rm{CP}}}} - l'} \right)}_M}} \right) ,
\end{align}
where ${\Xi _N}(x) \buildrel \Delta \over = \frac{1}{N}\sum\limits_{i = 0}^{N - 1} {{e^{ - {\bar\jmath}2\pi \frac{x}{N}i}}} $, which also shows up after (\ref{equ2-22}).

Next, by substituting (\ref{equA-2}) into (\ref{equA-1}), ${{Y^{{\rm{DD}}}}\left[ {k,l} \right]}$ can be written as	
\begin{align}\label{equA-3}
&{Y^{{\rm{DD}}}}\left[ {k,l} \right] \nonumber \\
&= \sum\limits_{i = 0}^{P - 1} \sum\limits_{{s_i} = 0}^{S - 1} \sum\limits_{k' = \left\lceil {{{ - N} \mathord{\left/
						{\vphantom {{ - N} 2}} \right.
						\kern-\nulldelimiterspace} 2}} \right\rceil }^{\left\lceil {{N \mathord{\left/
						{\vphantom {N 2}} \right.
						\kern-\nulldelimiterspace} 2}} \right\rceil  - 1}{X^{{\rm{DD}}}}\left[ {{{\left\langle {k - k'} \right\rangle }_N},{{\left( {l - {l_i}} \right)}_M}} \right]  \nonumber \\
&\ \times\!{h_{{s_i}}}{\Xi _N}\!\left( {k' - {k_{{s_i}}} - {{\tilde k}_{{s_i}}}} \right) \!  {e^{{\bar\jmath}2\pi \frac{{\left( {l - {l_i} + {M_{{\rm{CP}}}}} \right)\left( {{k_{{s_i}}} + {{\tilde k}_{{s_i}}}} \right)}}{{\left( {M + {M_{\rm{CP}}}} \right)N}}}} \!\!\!+\! {Z^{{\rm{DD}}}}\left[ {k,l} \right]  \nonumber \\
&=\frac{1}{N}\sum\limits_{i = 0}^{P - 1}  \sum\limits_{k' = \left\lceil { - {N \mathord{\left/
				{\vphantom {N 2}} \right.
				\kern-\nulldelimiterspace} 2}} \right\rceil }^{\left\lceil {{N \mathord{\left/
				{\vphantom {N 2}} \right.
				\kern-\nulldelimiterspace} 2}} \right\rceil  - 1}  \sum\limits_{j = 0}^{N - 1}  \sum\limits_{{s_i} = 0}^{S - 1}{X^{{\rm{DD}}}}\left[ {{{\left\langle {k - k'} \right\rangle }_N},{{\left( {l - {l_i}} \right)}_M}} \right]  \nonumber \\
&\ \times\!{h_{{M_{{\rm{CP}}}} + j\left( {M + {M_{{\rm{CP}}}}} \right),{l_i}}^{{s_i}}{e^{{\bar\jmath}2\pi \frac{{l\left( {{k_{{s_i}}} + {{\tilde k}_{{s_i}}}} \right)}}{{\left( {M + {M_{{\rm{CP}}}}} \right)N}}}}}  {e^{ - {\bar\jmath}2\pi \frac{{k'j}}{N}}} \! +\! {Z^{{\rm{DD}}}}\left[ {k,l} \right] ,	
\end{align}
which completes the proof.

\section*{Acknowledgment}
The authors would like to thank the associate editor and the anonymous reviewers for their helpful comments and suggestions.

\bibliographystyle{IEEEtran}
\bibliography{massive_MIMO_OTFS_final}

\end{document}